\documentclass[%
reprint,
superscriptaddress,
showpacs,
nofootinbib, nobibnotes, 
 amsmath,
 amssymb,
 aps,
 prl
]{revtex4-1}

\usepackage{natbib}
\usepackage{microtype}
\usepackage{blindtext}
\usepackage{verbatim}
\usepackage{float}
\usepackage[inline]{enumitem}
\usepackage{multirow}
 \usepackage{bbm}
\usepackage{dcolumn}
\usepackage{bm}
\usepackage{color, soul, colortbl} 
\usepackage[colorlinks=true,
						linkcolor=blue,
						urlcolor=blue,
						citecolor=blue,
						bookmarks=true,
						pdfborder={0 0 0}]{hyperref}
\usepackage{bbm}
\usepackage{booktabs}

\usepackage{xcolor}
\usepackage{mdframed} 

\usepackage{graphicx,epsfig}
\usepackage{subfigure}
\usepackage{hyperref}
\usepackage{natbib}
\usepackage{amsmath,amsfonts}
\usepackage{xfrac}
\usepackage[overload]{empheq}
\usepackage{mathtools}
\usepackage{cases} 
\usepackage{cancel} 

\begin{document}
\title{On the need to move from a single indicator to a multi-dimensional framework to measure accessibility to urban green}

\author{Alice Battiston}
\affiliation{University of Turin, Via Giuseppe Verdi, 8, 10124 Torino TO, Italy}
\affiliation{Barcelona Supercomputing Center, Plaça d'Eusebi Güell, 1-3, 08034 Barcelona, Spain}
\author{Rossano Schifanella}
\affiliation{University of Turin, Via Giuseppe Verdi, 8, 10124 Torino TO, Italy}
\affiliation{ISI Foundation, Via Chisola 5, 10126 Torino TO, Italy}


\maketitle

With the recent expansion of urban greening interventions, the definition of spatial indicators to measure the provision of urban greenery has become of pivotal importance in informing the policy-design process. By analyzing the stability of the population and area rankings induced by several indicators of green accessibility for over 1,000 cities worldwide, we investigate the extent to which the use of a single metric provides a reliable assessment of green accessibility in a city. The results suggest that, due to the complex interaction between the spatial distribution of greenspaces in an urban center and its population distribution, the use of a single indicator might lead to insufficient discrimination across areas or subgroups of the population, even when focusing on one form of green accessibility. From a policy perspective, this indicates the need to switch toward a multi-dimensional framework that is able to organically evaluate a range of indicators at once.

\section{Introduction}\label{sec:introduction}
As the share of the worldwide population living in cities is forecast to rise up to 68\% by 2050 \cite{united20182018}, urban greening interventions (UGIs) and nature-based solutions (NBSs) are increasingly relied upon to improve health outcomes and the well-being of urban communities as well as to mitigate the environmental footprint of cities \cite{hunter2017urban, bauduceau2015towards}. On the one hand, the use of Public Green Areas (PGAs) by local communities -for physical activity, leisure, and social exchange- has been associated with healthier lifestyles and increased social cohesion \cite{europe2017urban}. In this regard, studies have shown that living next to a PGA leads to reduced mortality rates, lower risk of cardiovascular diseases, and improved mental health and cognitive functions \cite{kondo2018urban, callaghan2021impact}. On the other hand, PGAs are effective solutions to pressing environmental challenges, providing biodiversity support and carbon storage but also acting as soil protectors and temperature regulators \cite{goddard2010scaling, shafique2020overview, shishegar2014impact, massaro2023spatial}.  

The multi-faceted benefits of urban green found definitive recognition in the New Urban Agenda, adopted at United Nations Conference Habitat III in 2016 \cite{agenda2016habitat} and in the \href{https://sdgs.un.org/goals/goal11}{ United Nations (UN) Sustainable Development Goals (SDG)}. Goal 11.7, in particular, calls for the universal provision of safe, inclusive, accessible, green, and public spaces for all demographics and, specifically, for the most vulnerable. Based on this guiding principle, health organizations, local authorities, and other institutional bodies have defined a range of green targets for cities to be met. For instance, the World Health Organization (WHO) recommends access to at least 0.5-1 ha of public green within 300m of residential locations for urban residents \cite{europe2017urban}. The wide spectrum of benefits associated with exposure to nature has also been encoded into multi-level targets, such as the one set by Natural England (2010) \cite{england2010nature} or by the city council of Berlin, Germany \href{https://www.berlin.de/umweltatlas/_assets/nutzung/oeffentliche-gruenanlagen/de-texte/kc605_2020.pdf}{[Berlin Senate Department for Urban Development and Housing, 2020)]}. The former is a 6-level target, with each level setting green provision requirements for increasing distances from residential locations. The latter articulates in three components: a short-range target (access to a PGA of 0.5ha within 500m or 5 to 10 minutes walking), a medium-range target (access to a PGA of 10ha 1 to 1.5 km), and a per-person target (access to at least 6m2 of smaller and 7m2 of larger green areas per person). In a similar fashion, the recently proposed 3-30-300 paradigm addresses the need for urban green to percolate into the life of urban residents at several levels, requiring that three trees must be visible from every home, every neighborhood must have a 30\% tree canopy cover and every home must have a greenspace within 300 meters \cite{konijnendijk20213}.

The proliferation of indexes and targets also reveals a progressive shift of the urban planning paradigm towards more data-driven policy design processes as a pathway to healthier and more sustainable cities. In this regard, the development of spatially-resolved indicators is seen as a first milestone to monitor progress towards specific goals but also to develop future data-driven urban policies, according to the principle \textit{what gets measured, gets done} \cite{giles2022creating, cohen2017systematic}.  

\begin{figure*}[t]
\includegraphics[width=\linewidth]{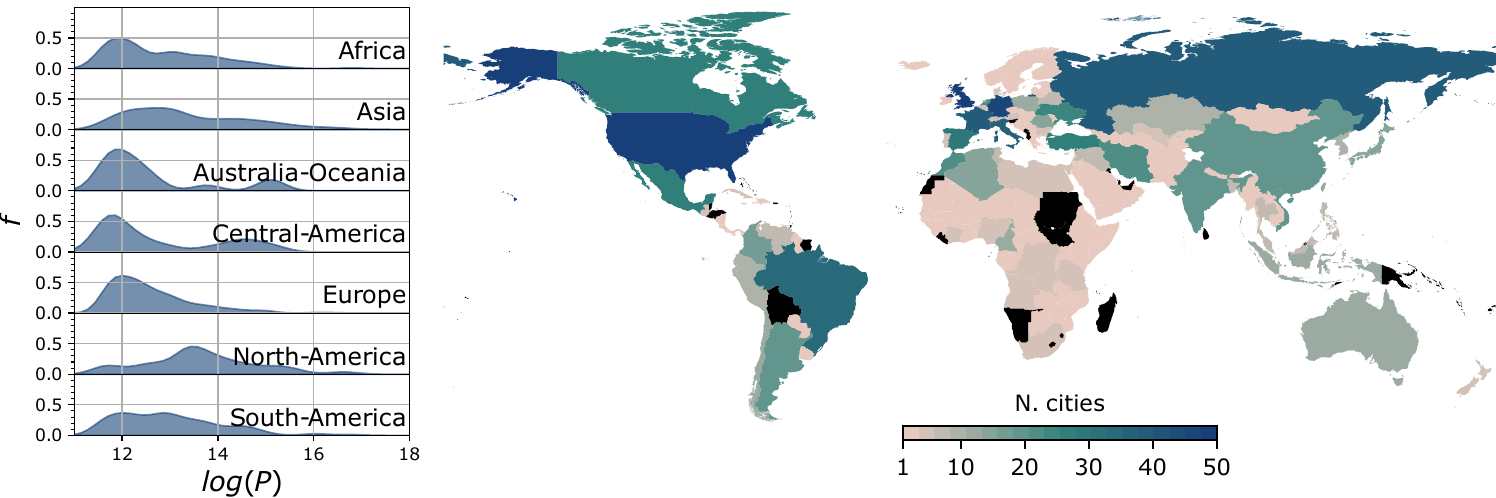}
\caption{\textbf{Geographical coverage of the study.} The map displays the number of cities included in the study, for each country. Black indicates that no city was included for the corresponding country. The density plot on the left reports the size (measured as log-population) distribution of cities included in the sample for each macro-area. Cities in the Russian Federation have all been attributed to Europe in this panel.}
\label{GeoCoverage}
\end{figure*}

Despite this, there is not yet an established and universally adopted framework to measure accessibility to urban green. Depending on the specific application and the available data sources, scholars in this domain have preferred one indicator to another, ranging from the minimum distance to the nearest park --using land use data from administrative sources, open crowd-sourced geodatabases (such as OpenStreetMap (OSM)) or processed satellite images \cite{rigolon2018inequities, oh2007assessing, zhang2021measuring, rahman2018analyzing, giuliani2021modelling, boeing2022using}-- to metrics evaluating the total green exposure from satellite data on land cover or on green intensity measured through the normalized difference vegetation index (NDVI) \cite{chen2022contrasting, song2020does, barboza2021green}. 
Moreover, unlike average city-level metrics, the definition of spatial indicators poses additional computational and methodological challenges as it requires considering the interplay between the spatial distribution of the population and greenspaces within a city, but also the walkable catchment of each sub-area in the city as resulting from the topology of its street-network. A first large-scale characterization of green exposure to partially account for this interplay is provided in the Global Human Settlement Urban Center Database (GHS-UCDB) \cite{florczyk2019}, which includes a measure of the \textit{generalized potential access to green areas}. This metric captures the \textit{immediate} exposure to green, as residents are considered to be exposed exclusively to the green available within their residential cell, regardless of the characteristics of nearby areas. A recent study acknowledges the importance of measuring short-range, as well as long-range metrics \cite{chen2022contrasting}. The authors compute green exposure for buffered regions with three progressively increasing radii around residential locations, for cities in the Global South and Global North. Despite the recent progress in the characterization of the interplay between population and greenspaces, considerations on walkable catchment areas are still mostly neglected in the majority of large-scale studies due to their computational complexity (conversely walkable catchments are typically considered in single-city settings, such as \cite{oh2007assessing, zhang2021measuring, rahman2018analyzing, giuliani2021modelling}). A first attempt in this sense is provided in \citeauthor{boeing2022using}, for PGAs within 500 meters of residential locations.

In this study, we argue that a comprehensive assessment of green provision in urban environments cannot overlook the inherently multi-dimensional nature of green accessibility, well-exemplified by the existence of multi-level targets. To this scope, we propose a framework to operationalize the measurement of three families of green accessibility indicators (minimum distance, exposure, and per-person) under many parametrizations (e.g. type of greenery, the minimum size of the PGA of interest, and/or the time budget). The aim is to measure --within a unified framework-- all those forms of green accessibility that contribute to the definition of the nature-citizen experience in urban environments.
The performance of the multi-indicator framework compared to a single-indicator setting is then evaluated along two dimensions. First, we assess the stability of each class of green accessibility indicators to small changes in its underlying parameters to unveil the impact that a fixed parametrization has on the induced ranking of areas and residents within a city. Second, we study the degree of overlap of targets set by institutional bodies to investigate to what extent the accessibility pictures emerging from the various metrics are interchangeable.
This study covers over 1,040 cities in 145 countries and the database built for this analysis is made fully accessible to policy-makers and the general audience through a \href{http://atgreen.hpc4ai.unito.it/}{dedicated interactive web platform}. The platform provides a range of functionalities, from the exploration of standard green accessibility targets to the generation of new indicators with customized settings, under the current green infrastructure of a city or after the implementation of a user-defined UGI.

\section{Results}\label{sec:results}
\subsection{Families of accessibility indexes}
\begin{figure}[t]
\includegraphics[width=\linewidth]{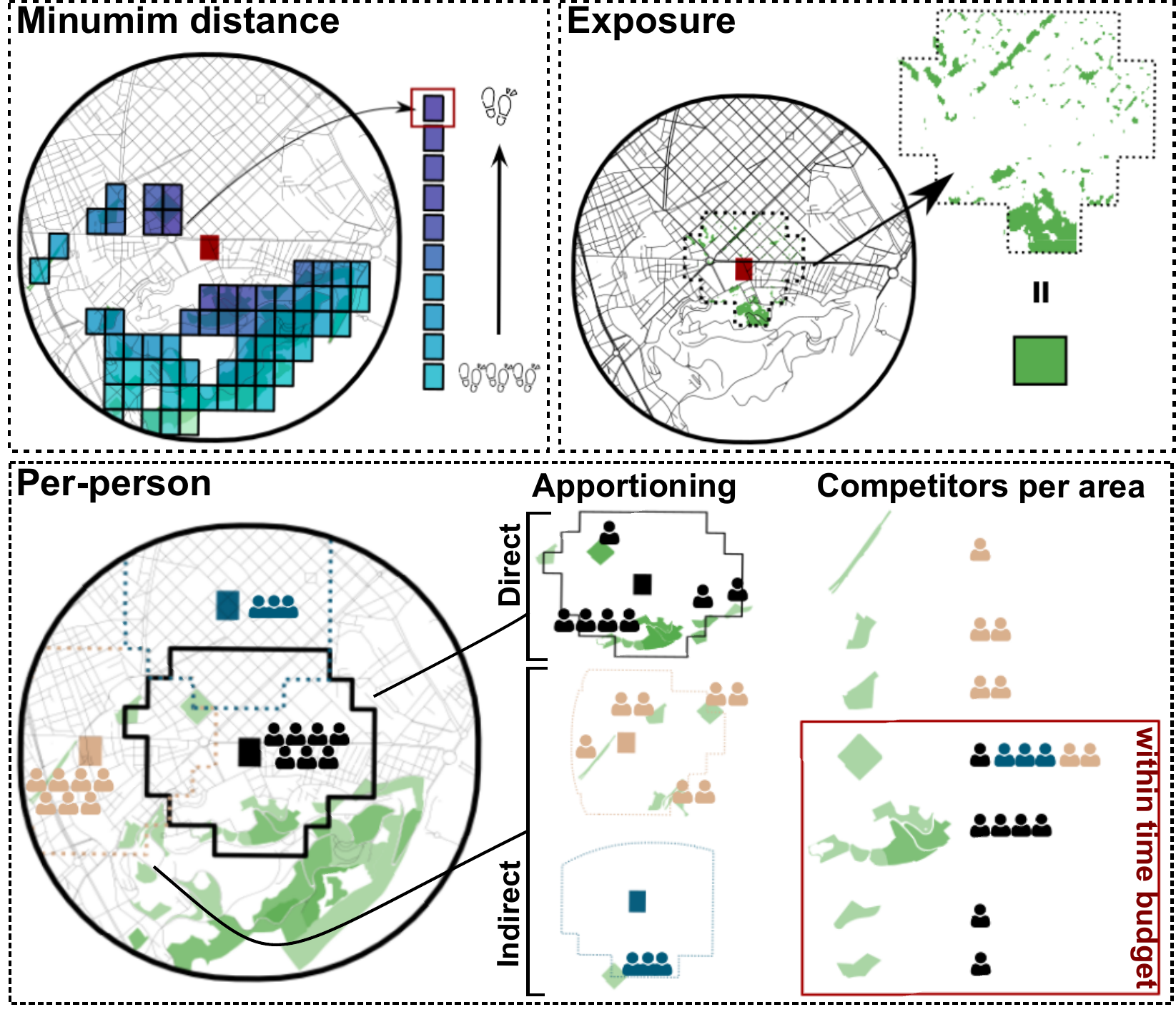}
\caption{\textbf{Graphical representation of the three families of accessibility indexes.} The panel displays a schematic representation of the proposed three families of green accessibility indicators. \textit{Minimum distance}: The indicator measures the walking distance in minutes to the closest cell with a PGAs with selected characteristics in terms of size and type of green, for each residential cell. In the top-left panel, the red cell is the cell of interest. The remaining cells are nearby cells equipped with a PGA, whose color is proportional to the distance to the cell of interest. \textit{Exposure}: The indicator measures the cumulative size of green features (in hectares) available within a walking time budget from a residential cell. In the top-right panel, the red cell is the cell of interest. The area within the walking time budget from the cell of interest is depicted with a dotted line. \textit{Per-person}: This indicator is computed in two steps. In the first step, all residents in the urban center are apportioned to PGAs -- within the corresponding walking time budget -- proportionally to the size of the PGA. In the second step, for each residential cell, the per-person index is computed as the ratio between the size of PGAs within the walking time budget from the cell of interest and the total number of residents apportioned to these areas --irrespective of their residential location. In the bottom panel, the cell of interest, its residents, and the area within the corresponding walking time budget are depicted in black. The same information for competitor users from other residential cells is depicted in pink and blue. A formal definition of each family of indicators is provided in the \hyperref[MatMeth]{Materials \& Methods}.}
\label{FigIndicators}
\end{figure}
Following the literature and policy debate on the topic, we built a framework to measure three families of indicators to characterize access and exposure to urban greenery for over 1,040 cities worldwide (Figure \ref{GeoCoverage}): 
\begin{itemize}
\item \textit{Minimum distance indicator} - This indicator measures the walking distance in minutes from a residential location to the closest PGA.
\item \textit{Exposure indicator} - This indicator measures the total exposure in hectares (ha) to urban green from a residential location within a certain (walking) time budget.  Compared to the minimum distance index, where we only focus on \textit{accessible} and \textit{public} green, the scope of the exposure indicator is to measure the cumulative availability of green features around a residential location -- irrespective of their use and degree of accessibility. This means, for example, that for this indicator we do not distinguish between public and private green, but also that we extend the definition of the green of interest to include other green features that populate an urban environment, such as a line of trees alongside a road. From the data viewpoint, this different perspective is mirrored in the change of the data source used to identify green elements (from OSM  to the 2020 World Cover database of the European Space Agency). 
\item \textit{Per-person indicator} - This indicator measures the per-person availability in square meters of PGAs within a (walking) time budget around a residential location. Unlike the previous two indicators, which are agnostic to population density, this index incorporates the notion that the use of PGAs for specific activities is competitive. Therefore, the level of public green that is available to a resident does not depend only on the total green provision but also on the cumulative number of people living within the service area of each PGA.
\end{itemize}
All families of indicators can be parameterized according to the minimum size of the PGAs/green features of interest, the type of greenery and the time budget (when applicable).
A schematic representation of each family is provided in Figure \ref{FigIndicators} and a formal definition is provided in the \hyperref[MatMeth]{Materials \& Methods}. 

\subsection{Stability of green accessibility indicators}
\begin{figure*}[t]
\includegraphics[width=\linewidth]{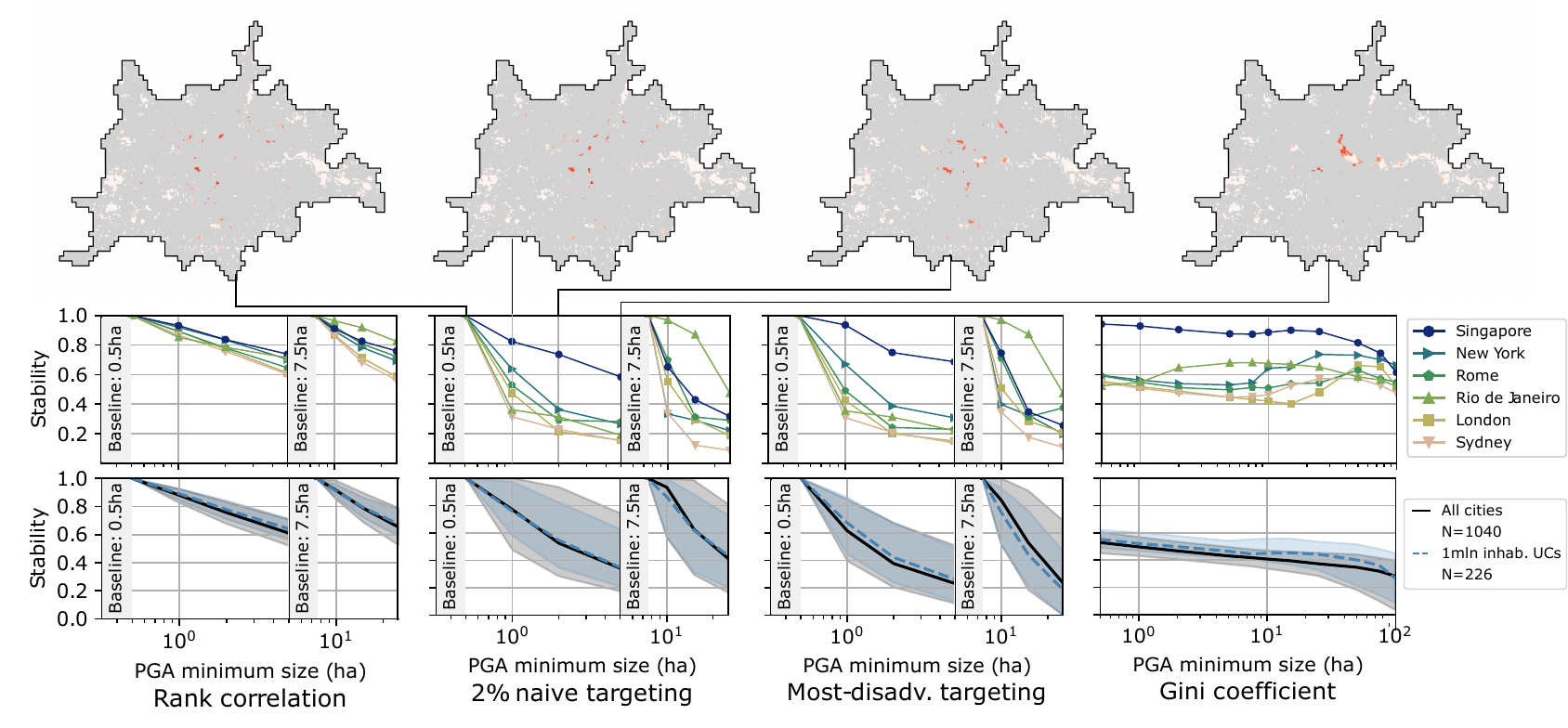}
\caption{\textbf{Stability of minimum distance index with respect to the minimum size of the PGA.} The top row depicts in red target areas in London (UK) according to the \textit{2\% naive targeting approach} for the minimum distance index with increasing minimum PGA size (from left to right: 0.5 ha, 1 ha, 2 ha, and 5 ha). The intensity of the red is proportional to the number of residents in the cell. The middle row depicts the level of stability of the minimum distance index to different parameterizations and according to different stability metrics, for six cities across all continents. The \textit{Most-disadvantaged targeting} targets residents performing worse than 3 times the mean citizen. For the \textit{Rank correlation}, \textit{2\% naive targeting} and the \textit{Most-disadvantaged targeting} the comparison is provided with respect to the parametrization with minimum size equal to 0.5ha for minimum sizes up to 5 ha and to 7.5ha for larger minimum sizes. For the \textit{Gini index}, the chart reports the value of the index under several parametrizations. The bottom row reports the median value (solid line) and the IQR (shaded area) of the stability metrics for all cities in our sample (black) and for cities with more than 1 million inhabitants (blue). A formal definition of each stability metric is provided in the \hyperref[MatMeth]{Materials \& Methods}. }
\label{Stability_MD}
\end{figure*}

\begin{figure}[t]
\includegraphics[width=\linewidth]{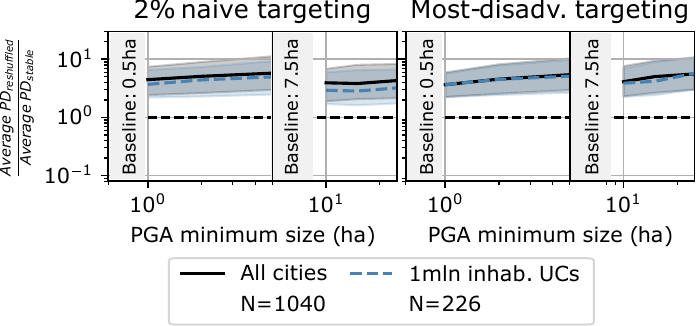}
\caption{\textbf{Population density of reshuffled and stable targeted areas.} The panel depicts the ratio of the average population density ($PD$) of the areas associated with the reshuffled targeted population to the average $PD$ of areas linked to the stable targeted population, under two targeting strategies and a range of alternative minimum PGA size for the minimum distance indicator. The line is the median across all cities (in black) and cities with more than 1 million inhabitants (blue); the IQR for both cases is depicted as a shaded area.}
\label{PDensity_MD}
\end{figure}

Institutional targets and guiding principles are often ambiguous in terms of the exact parametrization of an indicator (as an example, see \cite{WHO}, \cite{england2010nature}). In this section, we evaluate the sensitivity of each family of indicators to changes in its parameterization along three dimensions and define, whenever appropriate, a corresponding stability metric. The three dimensions are:
\begin{enumerate}
\item the ranking of sub-areas in the urban center according to their level of green accessibility. The stability of the rankings induced by two alternative parameterizations is measured through the Kendall rank correlation coefficient \cite{kendall1948rank}.
\item the ranking of residents of the urban center according to their level of green accessibility. This dimension differs from the previous one due to the non-uniform distribution of the population over the urban area. Rather than evaluating the entire ranking, here we adopt a policy perspective and assess the stability of the subgroup of the population at the bottom of the ranking, as we might expect this group to be targeted for potential interventions. To this scope, we propose two fictional targeting strategies. With the first strategy, labeled \textit{naive targeting}, we target the $y\%$ worst performing population, irrespective of the actual performance; with the second strategy, labeled \textit{most-disadvantaged targeting}, we target subgroups of the population with low performance in either absolute term or relative to the rest of the population (see \hyperref[MatMeth]{Materials \& Methods}). We then define the stability of an indicator to any two parametrizations (baseline vs alternative), under a targeting strategy, as the share of the overlapping targeted population (see \hyperref[MatMeth]{Materials \& Methods}).  For a stability level $s$, the proportion of the target population under one parameterization that would not be targeted under the alternative one (hereafter: reshuffled target population) is given on average by $\frac{(1-s)}{(1+s)}$.
\item the observed level of inequality in the urban center under the specific parameterization of the index, measured through a weighted Gini Index \cite{gini1936measure}.
\end{enumerate}
Figure \ref{Stability_MD} displays the stability of the minimum distance index against changes in the minimum size of the PGA, for all cities in our sample (mean and inter-quartile range (IQR)), for cities larger than 1 million inhabitants (mean and IQR) and for selected large urban centers across all continents. 
The stability of parametrizations with a minimum PGA size of up to 5 ha is assessed against a baseline parametrization with a minimum PGA size of 0.5 ha. The stability of larger parametrizations is assessed against a baseline with a minimum PGA size of 7.5 ha.
Across all stability metrics, we observe a progressive decrease in the median level of stability across cities as we increase the minimum size of the PGA, with no substantial differences among large cities with more than 1 million inhabitants and all cities. Higher stability values measured at the area level (Figure \ref{Stability_MD}-\textit{Rank correlation}) hide relevant reshuffling at the bottom of the population ranking. Under the \textit{2\% naive targeting strategy} the median stability level across all cities in our sample stands at 0.77 [IQR: 0.48-1] and 0.53 [IQR:0.28-0.94] for a shift in the parameters from 0.5 ha to 1 ha and to 2 ha respectively. This corresponds to a reshuffling in the target population of 12\% [IQR: 35\%-0\%] and 30\% [IQR: 56\%-3\%]. On the one hand, these results highlight large variability in the stability level of the index depending on the specific green configuration of the urban center; on the other hand, they also suggest worryingly levels of instability, particularly for selected large cities. Indeed, for the same parameterizations (minimum PGA size of 0.5 ha and 1 ha), the stability levels of the cities of Sydney (AUS), Rio de Janeiro (BRA), and London (UK) stand at 0.32, 0.36, and 0.47 respectively, corresponding to mean population reshuffling levels of 51\%, 47\%, and 36\%. Similar results emerge by using the more restrictive \textit{most-disadvantaged targeting strategy}, which targets populations with low accessibility levels more tightly, and by setting different targeting levels for the naive targeting approach (see Supplemental Material (SM)). A visual assessment of the changes in the targeted population for the city of London (UK) is provided in the maps in the top row of Figure \ref{Stability_MD}, which display -- in a scale of reds --targeted areas using the \textit{2\% naive targeting strategy} under 5 parameterizations, with the intensity of the palette being proportional to the number of people living in the area. From the graphical comparison, we observe that:
\begin{enumerate}
\item For small changes in the parameter (e.g. from 0.5 ha to 1 ha), most of the stable population (i.e., populations targeted under both parameterizations) is concentrated in low-density areas. While some (but fewer) higher-density areas are targeted in both scenarios, they tend not to overlap. This entails an additional challenge from the viewpoint of policymakers, as stable areas might not be sufficiently populated to be meaningful areas of intervention. It is worth noticing that this is not a peculiarity of London only but holds for the vast majority of cities: in Figure \ref{PDensity_MD}, we show that the ratio of the population density of the areas associated with the reshuffled targeted population to the population density of areas linked to the stable targeted population is consistently above 1.
\item The degree of clustering of targeted areas increases along with the minimum size of the PGAs. This is due to the physical and geographical constraints posed by larger PGAs, which are typically fewer and less scattered around the city than smaller PGAs.
\end{enumerate}

Finally, we explore the impact that the parametrization has on the level of inequality within a city as measured through a weighted Gini index. Interestingly, we do not observe any common pattern across cities. While some cities experience decreasing levels of inequality (e.g. Singapore (SGP)) as we increase the minimum size of the PGAs, others show increasing levels (New York (USA)) or U-shaped patterns (Sydney (AUS) and London (UK)). While this is likely to be a result of the interplay between the size composition of PGAs within a city and its spatial distribution, we do not assess here the existence of specific regularities. Similar figures for the stability of the exposure and per-person indicators against the time budget (for both) and the minimum size of the PGAs (for the latter only) are provided in the SM as well as the sensitivity analysis to the targeting strategies for all indicators. For the exposure indicator, we observe a median \textit{Rank correlation} for cities in our sample of 0.69 [IQR: 0.63-0.74] for a shift in the time budget from 5 to 10 minutes and median stability levels of the \textit{2\% targeting strategy} of 0.38 [IQR: 0.23-0.63] for a similar change in the time budget. See the additional material for detailed stability metrics across our entire sample.

\subsection{Stability of targets set by selected institutional bodies.}

\begin{figure}[h]
\includegraphics[width=\linewidth]{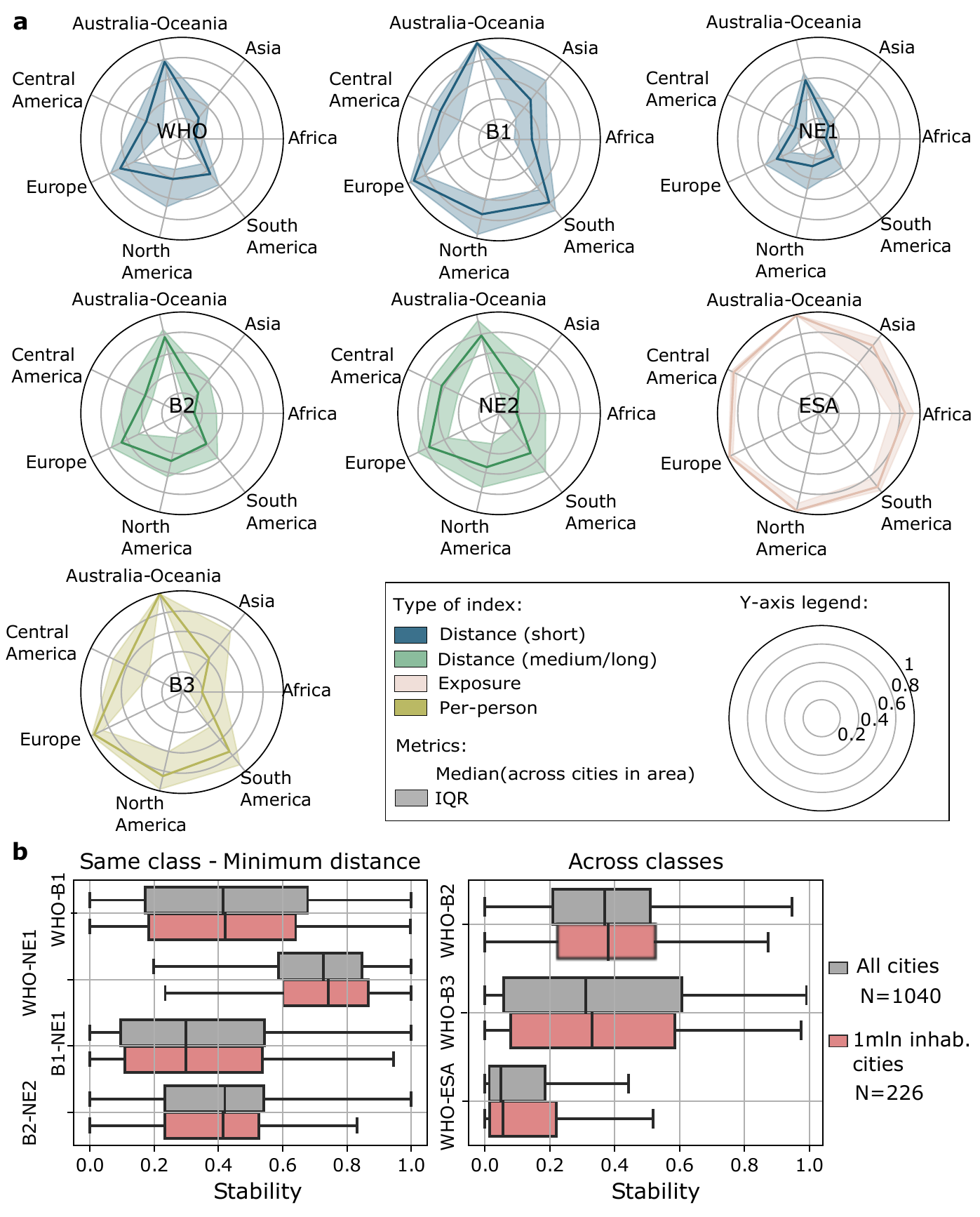}
\caption{\textbf{Indexes and targets proposed by institutional bodies.} a) For each institutional target (see Table \ref{tab:indicesDefinitions}), the radial plot depicts the median (solid line) and IQR (shaded area) across cities in each geographical macro-area of the proportion of the population satisfying the target. The color of the plot reflects the class of the target. Cities in the Russian Federation have been attributed to Europe. b) The box plots depict the cross-indexes level of stability in the population non-satisfying the corresponding institutional targets, for all cities in the sample (black) and for cities with more than 1 million inhabitants (red). }
\label{Fixedindices}
\end{figure}

\begin{table}[h]
\centering
\begin{tabular}{llllll}
\hline
\textbf{Index} & \textbf{Type} & \textbf{Size} & \textbf{Distance} & \textbf{Target} & \textbf{Data} \\ \hline
\textbf{WHO} & Min. distance & 0.5 ha & 5 mins  & 5 mins  & OSM  \\ 
\textbf{B1}  & Min. distance & 0.5 ha & 10 mins & 10 mins & OSM   \\
\textbf{N1}  &Min. distance & 2 ha   & 5 mins  & 5 mins  & OSM   \\
\textbf{B2}  & Min. distance & 10 ha  & 15 mins & 15 mins & OSM   \\
\textbf{N2}  & Min. distance & 20 ha  & 25 mins & 25 mins & OSM   \\
\textbf{B3}  & Per-person       & 0.5 ha & 15 mins & 6 mq$^{2}$  & OSM   \\
\textbf{ESA} & Exposure         & 100 mq$^{2}$      & 5 mins  & 0.5 ha  & ESA   \\ \hline
\end{tabular}

\caption{\textbf{Operational definition of indexes and targets proposed by institutional bodies.} The proposed indexes are inspired by targets set by local authorities and public-health bodies worldwide. An index is an underlying metric used to measure whether an area satisfies the corresponding target. For instance, the WHO index measures the walking distance in minutes from the closest PGA of at least 0.5 hectares, and an area satisfies the WHO target if the walking distance to the closest PGA of at least 0.5 hectares is no more than 5 minutes. Column $Size$ refers to the minimum size of greenspaces extracted for the computation of the index. Column $Data$ refers to the data source used for the extraction of greenspaces. It should be noted that the ESA index is the only indicator using green data from the WC-ESA 2020.}
\label{tab:indicesDefinitions}
\end{table}

In the previous section, we studied the stability of green accessibility indicators under several parameterizations. Here, we focus on seven green accessibility indicators inspired by institutional targets, defined by public-health bodies and local authorities worldwide. An operational definition of each target is provided in Table \ref{tab:indicesDefinitions}. 
The existence of well-defined green targets naturally induces a metric to measure the performance of cities in terms of green accessibility, i.e. measuring the proportion of the inhabitants of a city that satisfies the prescribed target. By coupling information on the green indexes and the population density, we measure the proportion of the population satisfying each target in each city. Figure \ref{Fixedindices}a summarizes the performance of cities in our sample for each target grouped by geographical area.  The results display a marked geographical pattern, in line with what was observed in analogous studies \cite{chen2022contrasting, han2023inequality}. The multiplication of green indexes reflects a recent spike in interest by local authorities and public-health bodies for the development of greener urban environments as well as the recognition of the wide range of benefits associated with exposure to nature. At the same time, with the exception of the target set by the WHO, this proliferation also manifests the existence of concurrent authorities operating at the same level and setting their own goals independently. In this section, we assess the extent to which the accessibility pictures emerging from these indexes are interchangeable and show how the use of one index/target alone is insufficient to capture the full complexity of green accessibility, thus restating the need to use an organic framework measuring several indexes at once. 
Similarly to the previous section, we quantify the degree of disagreement between any pair of two indexes for each city through the stability of the population not satisfying the target proposed by the institutional body. 
Figure \ref{Fixedindices}b shows the degree of stability across any two pairs of indexes, for indicators within the same class or across different classes. Unsurprisingly, we observe greater stability for indexes within the same class, than across indexes belonging to different classes. Interestingly, the lowest stability is observed between the WHO index and the ESA, reflecting the different types of green features incorporated in the two indexes (in particular, the incorporation of green elements different from parks, grassland, and forests in the latter). 

\section{Discussion}\label{sec:discussion}
While most of the recent literature on accessibility to urban greenery focused on comparing cities or areas within a city based on one specific index, this study aimed at stressing the importance of considering the inherently multi-dimensional nature of green accessibility when evaluating the provision of green in an urban environment. To this scope, we assessed the degree of similarity of the accessibility pictures emerging from indicators within the same family, but under different parametrizations, and across different families. The results of the study indicate substantial differences in the ranking of areas and populations induced by different parametrizations of the same family of indicators or by indicators across two families. From the policy perspective, this means that an assessment based on one specific set of parameters may be insufficient to identify subgroups of the population that are missing out and highlights the need to switch towards a multi-indicator evaluation.

The study used a set of fictional targeting strategies for the identification of subgroups of the population within a city with little access to nature. It is worth noting that we do not expect these simplistic strategies alone to be representative of a realistic policy-design process. On the one hand, there can be built-environment or financial constraints, so that even severely under-performing areas may not be viable targets for UGIs due, for instance, to the lack of suitable patches for intervention or low economic returns; on the other hand, there might be reasons to prefer the targeting of specific demographic groups, as for instance elderly or children whose ability to move around the city might be more limited.  At the same time, it is worth remembering that the complexity of the design of UGIs goes well beyond the identification of the target populations and areas, requiring measures to deal with so-called green gentrification phenomena \cite{anguelovski2022green, rigolon2020green, anguelovski2018assessing, maia2020hidden, garcia2021urban, rigolon2023green} and to increase citizens' engagement with the UGI through co-creation processes. 
While we devoted a large part of our time during this research activity to the cleaning and processing of the data to ensure the best possible standards, the main limitation of our work concerns the completeness of the mapping of green features in OSM. To limit the impact of this potential data bias, we undertook a series of filtering and checks to assess the quality of the OSM data in each urban center (see SM), resulting in more than halving the initial sample of cities (from around 2,500 to 1,040). However, the accessibility metrics that we measure are intrinsically dependent on the quality of these data so that low green feature quality would necessarily result in biased indicators. In an effort of transparency and to facilitate the identification of these biases, we make all our data (including the raw data) easily navigable to the public with our interactive platform. Nonetheless, we deem the impact of this issue on the stability metrics performed in this study to be limited as the comparison is always provided within the same city rather than across cities. As such, this limitation does not undermine the main take-home message of this study. 
Another limitation of the work concerns the definition of PGAs for the minimum distance and the per-person indicators. Our definition of urban green is fully reliant on the mapping of the area in OSM and as such on the assessment of the mapper(s) on the correct tag. In the presence of heterogeneous standards for the mapping of PGAs, it is possible that the characteristic of a PGA in terms of the level of green, type of services, and characteristics of the vegetation varies largely depending on the country or the climate zone of the urban center. Once again, the interactive platform provides an initial attempt to control for this variability by giving the user the option to customize the type of green to incorporate in the index, from a more restrictive definition to a more extensive one.

\section{Materials \& Methods} \label{MatMeth}
\subsection{Definition of urban centers}
Urban centers (UCs) --or cities, here used interchangeably-- were defined according to the boundaries in the Urban Centre Database of the Global Human Settlement 2015, revised version R2019A (GHS-UCDB) \cite{florczyk2019}. UCs in the GHS-UCDB are not based on administrative entities but on specific cut-off values on the resident population and the built-up surface share in a 1x1 km global uniform grid. Out of 13,000 urban centers recorded in the database, we retained the most populated 50 UCs per country (a country is identified by the internationally recognized three-letter ISO code), provided that they had at least 100,000 inhabitants. We further excluded UCs for which the quality of the OpenStreetMap (OSM) data \cite{OpenStreetMap} was deemed insufficient according to the procedure described in the section 
 \ref{SISEC_dataValidation} of the SI. The data validation was performed by comparing the green intensity appearing from OSM data (based on an extended definition of green) \cite{OpenStreetMap} to the green intensity from the  World Cover data 2020 from the European Space Agency (WC-ESA) \cite{zanaga_daniele_2021_5571936} and defining ad-hoc acceptance intervals for urban centers with different size and average green intensity based on the level of similarity observed for a set of reference cities. The final sample comprised 1,040 UCs across 145 countries. 

\subsection{Geographical units of analysis}
The geographical space of each UC was divided into a regular grid with a spatial resolution of 9 arc-seconds (geographic projection: WGS-84), mimicking the grid of the population layer of the Global Human Settlement 2015 (GHS-POP)\cite{ghs_pop}. Cells in the grid are the smallest geographical unit of analysis for this study, meaning that all metrics were measured at this geographical level and, whenever appropriate, aggregated into higher geographical levels (e.g. UC).

\subsection{Definition of PGAs and other green features}
In the manuscript, the terms \textit{Public Green Areas} (PGAs) and \textit{greenspaces} are used interchangeably to indicate accessible green areas of public use. The terms \textit{urban green}, \textit{green infrastructure} and \textit{greencoverage} instead are used to refer to all green features in an urban center, regardless of their use or their degree of accessibility. For each city, PGAs were extracted from OSM data following the pipeline described in the SM and reclassified into three classes: \textit{parks}, \textit{grass} and \textit{forests}. For each city, the \textit{green infrastructure} is extracted from the WC-ESA 2020 (codes: $10$, $20$ and $30$) following the pipeline described in the SM.

\subsection{Data sources}
For each UC, the accessibility metrics presented in this study were constructed by combining information from three data sources: 1 - the GHS-POP\cite{ghs_pop}, which provides granular worldwide population estimates (2015), based on census data and satellite information on built areas. 2 - OSM\cite{OpenStreetMap} (accessed in May 2022). OSM data were used to extract spatial information on the location of public green areas used for the minimum distance and the per-person indexes and to compute walking distances within any two cells in the city. 3 - the World Cover data 2020 from the European Space Agency (WC-ESA) \cite{zanaga_daniele_2021_5571936}, which provides the land cover inferred by Sentinel-1 and Sentinel-2 data at a 10 meters resolution. WC-ESA data were used as the source of information on green elements for the exposure index. In addition, the WC-ESA data were used in the validation of the sample of UCs described in section \ref{SISEC_dataValidation} of the SI. 

\subsection{Pre-processing of the data}
For each UC, the pre-processing of the data comprised five phases -- described in detail in Sections \ref{SISEC_DataCleaningProcessing}.A to \ref{SISEC_DataCleaningProcessing}.E of the SI. 
\begin{enumerate}
\item Extraction of the city boundary from the GHS-UCDB \cite{florczyk2019}.
\item Extraction of the population distribution from the GHS-POP \cite{ghs_pop}, by clipping the worldwide information with the boundary of the UC buffered with a 3 kilometers radius. The 3 kilometers buffer was applied to ensure the computation of the accessibility metrics was not biased for cells near the boundary of the UC. 
\item Extraction and processing of the OSM data on public green areas. This phase consisted of three sub-steps. 1-  extraction of a local \textit{osm.pbf} dumps from the \textit{osm.pbf} of the corresponding continent, by clipping the continent-wide file with the boundary of the UC buffered with a 3 kilometers radius. 2 - extraction of all relations and closed ways associated with public green areas from the local dump; 3 - remapping of the information on the public green areas to the base-grid used for the analysis. 
\item Extraction and processing of the data on green coverage from the WC-ESA 2020 \cite{zanaga_daniele_2021_5571936}. This phase consisted of two sub-steps. 1-  extraction of a local \textit{.tiff} from the global \textit{.tiff}, by clipping the worldwide file with the boundary of the UC buffered with a 3 kilometers radius, following the procedure described \href{https://esa-worldcover.org/en/data-access}{here}. 2 - remapping of the information on green coverage to the base-grid used for the analysis. 
\item Computation of the walking distance matrix for the centroids of the base-grid using the $foot$ profile of the Open Source Routing Machine (OSRM) engine \cite{luxen-vetter-2011} and the local \textit{osm.pbf} file extracted at point 2.
\end{enumerate}
The pre-processing of the data was performed in Python, JavaScript, and PostGIS.

\subsection{Accessibility indexes}
Following the literature on accessibility to nature in urban environments, we operationalized three classes of accessibility indexes -\textit{minimum distance}, \textit{exposure}, and \textit{per-person}. In what follow: $\mathcal{N}_{c}$ is an ordered set of $N$ elements representing the cells within the city boundary in the population grid of city $c$, $\mathcal{M}_{c}$ is an ordered set of $M$ elements representing the cells in the extended (3km-buffered) grid for city $c$, $D_{c}$ ($D^{+}_{c}$) is an ($N$$\times$$M$)-matrix (($M$$\times$$M$)-matrix) where each element $d_{i,j}$ ($d^{+}_{i,j}$) represents the walking distance in minutes between the $i$-th element of $\mathcal{N}_{c}$ ($\mathcal{M}_{c}$) and the $j$-th element of  $\mathcal{M}_{c}$.
\vspace{2mm}
\paragraph{Minimum distance} This class of indexes measures the walking distance (in minutes) to the closest public green area. The index can be parameterized according to the minimum size of the green area and the type of green (combinations of parks, forests, and grass). 
For each cell $i$ in the population grid $\mathcal{N}_{c}$, the index is defined as:
\begin{equation} 
md_{i, c}=min(D_{i, c} \odot gd_{c})
\end{equation}
i.e., the minimum of the Hadamard product between the $i$-th row of $D_{c}$ and the $M$-dimensional vector $gd_{c}$ taking value 0 or 1 to indicate the absence/presence of a green feature (with given characteristics in terms of size/type of green) in the corresponding cell of the extended grid $\mathcal{M}_{c}$.
\vspace{2mm}
\paragraph{Exposure} This class of indexes measures the overall size of public green available within $t$ minutes from a residential location. As for the previous class, the index can be parameterized according to the value of $t$ and the minimum size of the green area. 
For each cell $i$ in the population grid $\mathcal{N}_{c}$, the index is defined as the sum of the green intensities measured on cells that are no more than $t$-minutes away from the cell of origin. I.e.:
\begin{equation} 
exp_{i, t, c}=\mathbbm{1}_{(0, t]}(D_{i, c}) \times gi_{c}
\end{equation}
where $\mathbbm{1}_{(0, t]}(D_{i, c})$ is an indicator function mapping each element of the $M$-dimensional vector $D_{i, c}$ to 0 if $d_{i,j}$ is greater than $t$ and to 1 otherwise. $gi_{c}$ is an $M$-dimensional vector representing the size of the green features (with given characteristics in terms of size) in the corresponding cell of the extended grid $\mathcal{M}_{c}$.  
\vspace{2mm}
\paragraph{Per-person} This class of indexes measures the square meters per person of public green available within $t$ minutes from a residential location. As for the previous classes, the index can be parameterized according to the value of $t$, the minimum size of the green area and the type of green. 
For each cell $i$ in the population grid $\mathcal{N}_{c}$, the index is computed as:
\begin{equation}
pp_{i, t, c}=\mathbbm{1}_{(0, t]}(D_{i, c}) \times gpp_{t, c}
\end{equation}
where $gpp_{t, c}$ is an $M$-dimensional vector representing the squared meters of green available per-person in the corresponding cell of $\mathcal{M}_{c}$. More specifically, $gpp_{t, c}$ is computed by dividing the green available in each cell by the total confluent population. I.e.:
\begin{equation}
gpp_{t, c}=gi_{c}\oslash AP_{(t, c)}
\end{equation}
where $\oslash$ refers to element-wise division, AP is the $M$-dimensional vector whose element $ap_{j,t,c}$ equals the confluent population (for the time-threshold $t$) of the $j$-th element of $\mathcal{M}_{c}$.
The affluent population of cell $j$ is computed by assigning -- for each residential cell $i$ -- shares of the population to cells in $\mathcal{M}$ within the time-budget $t$ proportionally to the size of the available green in $j$. Formally, the confluent population of cell $j$ in $\mathcal{M}$ for time-budget $t$ is defined as: 
\begin{equation}
ap_{j, t, c}=\sum_{i \in \mathcal{M} }P_{i}\dfrac{\mathbbm{1}_{(0, t]}(d^{+}_{i,j,c})*gi_{j,c}}{\sum_{m \in \mathcal{M}}\mathbbm{1}_{(0, t]}(d^{+}_{i,m,c})*gi_{m,c}}
\end{equation}

\subsection{Stability metrics}
We evaluate the stability of each accessibility index to its parametrization through three metrics -- the Kendall rank correlation coefficient across areas, the proportion of stable targeted population according to a naive targeting approach, and the proportion of stable targeted population according to a most-disadvantaged targeting approach. A formal definition of each metric is provided below. 
\paragraph{Kendall rank correlation coefficient }
Let $R_{Ind(x)}[N]$ be the ranking induced by the accessibility index $Ind$ with parametrization $(x)$ on the set of cell $\mathcal{N}_{c}$ of the urban center $c$.
The Kendall rank correlation coefficient between two parametrizations $(1)$ and $(2)$ of index $Ind$ is given by:
\begin{equation}
\tau=1 - \dfrac{number \,  of \,  discordant \, pairs}{\binom{N}{2}}
\end{equation}
where two pairs $n_{1}$ and $n_{2}$ in $\mathcal{N}_{c}$ are said to be \textit{concordant} if either $R_{Ind(1)}[n_{1}]<R_{Ind(1)}[n_{2}]$ and $R_{Ind(2)}[n_{1}]<R_{Ind(2)}[n_{2}]$ or  $R_{Ind(1)}[n_{1}]>R_{Ind(1)}[n_{2}]$ and $R_{Ind(2)}[n_{1}]>R_{Ind(2)}[n_{2}]$, otherwise they are \textit{discordant}.

\paragraph{Naive targeting approach}
Let $Ind(x)[N]$ be the accessibility index $Ind$ with parametrization $(x)$ on the set of cell $\mathcal{N}_{c}$ of the urban center $c$. For the minimum distance index, let $t^\star_{x}(y)$ be  
\small
\begin{equation}
t^\star_{x}(y)=\min \Bigg\{t: \Bigg[\dfrac{\sum_{n \in \mathcal{N}_{c}} P_{n} (1-\mathbbm{1}_{[0,t)}(Ind_{x}[n])) }{\sum_{n \in \mathcal{N}_{c}} P_{n}}\Bigg]*100 \ge y \Bigg\}
\end{equation}
\normalsize
 For the exposure and per-person indexes, let $t^\star_{x}(y)$ be  
\begin{equation}
t^\star_{x}(y)=\min \Bigg\{t: \Bigg[\dfrac{\sum_{n \in \mathcal{N}_{c}} P_{n} \mathbbm{1}_{[0,t]}(Ind_{x}[n]) }{\sum_{n \in \mathcal{N}_{c}} P_{n}}\Bigg]*100 \ge y \Bigg\}
\end{equation}
Then, for any two parametrizations $(1)$ and $(2)$ of $Ind$, for the minimum distance index, we define the proportion of the stable targeted population under the $y\%$ naive targeting approach as the following weighted Jaccard Index:
\begin{equation}
S_{naive}(y)^{md}=\dfrac{\sum_{n \in \mathcal{N}_{c}} P_{n}(1-\mathbbm{1}_{1,t^\star,y,n})(1-\mathbbm{1}_{2,t^\star,y,n})}{\sum_{n \in \mathcal{N}_{c}}P_{n} (1 - \min[ \mathbbm{1}_{1,t^\star,y,n},\mathbbm{1}_{2,t^\star,y,n}])}
\end{equation}
where $\mathbbm{1}_{x,t^\star,y,n}=\mathbbm{1}_{[0,t_{x}^\star(y))}(Ind(x)[n])$.
For the exposure and per-person indexes: 
\begin{equation}
S_{naive}(y)^{exp,pp}=\dfrac{\sum_{n \in \mathcal{N}_{c}} P_{n}\mathbbm{1}_{1,t^\star,y,n}\mathbbm{1}_{2,t^\star,y,n}}{\sum_{n \in \mathcal{N}_{c}}P_{n} \max[\mathbbm{1}_{1,t^\star,y,n}, \mathbbm{1}_{2,t^\star,y,n}]}
\end{equation}
where $\mathbbm{1}_{x,t^\star,y,n}=\mathbbm{1}_{[0,t_{x}^\star(y)]}(Ind(x)[n])$.
It is noteworthy that given the presence of potential ties (some cells may have the same accessibility value) induced by Ind(x), the number of people belonging to the bottom $y\%$ of the induced ranking population might differ under $Ind(1)$ and $Ind(2)$.

\paragraph{Most-disadvantaged targeting approach}
The most-disadvantaged targeting approach is defined similarly to the naive targeting approach but instead of using a cutoff value that depends on the ranking of the population only (and as such agnostic with respect to the value of the index itself), here the cutoff value to define the target population depends on the value of the index. 

For the exposure and per-person indexes, we define the target group as the group of people with no exposure/per-person access at all under the parameterization $(x)$ of the index. As such, the proportion of the stable population in the target group under any two parameterizations $(1)$ and $(2)$ is defined as the resulting weighted Jaccard Index:
\small
\begin{equation}
S_{most-dis}^{exp,pp}=\dfrac{\sum_{n \in \mathcal{N}_{c}} P_{n}\mathbbm{1}_{[0]}(Ind(1)[n]) \mathbbm{1}_{[0]}(Ind(2)[n])}{\sum_{n \in \mathcal{N}_{c}}P_{n} \max [\mathbbm{1}_{[0]}(Ind(1)[n]), \mathbbm{1}_{[0]}(Ind(2)[n])]}
\end{equation}
\normalsize

For the minimum distance index, we define the most-disadvantaged target population as that population that performs $y$-times worse than the average behavior across all citizens. As such, letting $t^{m}_x$ be the average value of the index under the parametrization $(x)$, then: 
\small
\begin{equation}
S_{most-dis}^{md}(y)=\dfrac{\sum_{n \in \mathcal{N}_{c}} P_{n}(1-\mathbbm{1}_{1,t^{m}_1,y,n})(1-\mathbbm{1}_{2,t^{m}_2,y,n})}{\sum_{n \in \mathcal{N}_{c}}P_{n}(1-\min[ \mathbbm{1}_{1,t^{m}_1,y,n}, \mathbbm{1}_{2,t^{m}_2,y,n}])}
\end{equation}
\normalsize
where $\mathbbm{1}_{x,t^{m}_x,y,n}=\mathbbm{1}_{[0,yt^{m}_x)}(Ind(x)[n])$.



\section{Abbreviations}
\noindent GHS-POP: population layer of the Global Human Settlement 2015 \\
GHS-UCDB: Urban Centre Database of the
Global Human Settlement 2015, revised version R2019A \\
NBS: Nature-based Solution \\
NDVI: Normalized difference vegetation index \\
OSM: OpenStreetMap \\
OSRM: Open Source Routing Machine \\
PGA: Public Green Area \\
SM: Supplemental Material \\
UC: Urban Center \\
UGI: Urban Greening Intervention \\
WC-ESA: World Cover data
2020 from the European Space Agency \\
WHO: World Health Organization

\section{Availability of data and materials}
The Python code developed for this project is available at the following Github repository: \href{https://github.com/alibatti/atgreen.git}{https://github.com/alibatti/ATGreen.git}. The raw data are all publicly available. The processed data are available upon request and can be explored through the interactive platform at \href{http://atgreen.hpc4ai.unito.it/}{http://atgreen.hpc4ai.unito.it/}. The interactive web platform has five functionalities: \textit{EXPLORE}, \textit{MEASURE}, \textit{COMPARE}, \textit{CREATE} and \textit{DRAW}. For each urban center: \textit{EXPLORE}, enables the exploration of green areas, by size, type, and source;  MEASURE and \textit{COMPARE}, enable the exploration and the mutual comparison respectively of a selection of indexes and targets proposed by institutional bodies; \textit{CREATE} allows the user to create its own accessibility index with the desired selection of parameters; DRAW allows the user to explore the change in accessibility resulting from the addition of a new PGA in an area of her choice.

\section{Acknowledgements}
We thank Dr. Patricio Reyes, Dr. Fernando Cucchietti, and the rest of the Data Analytics and Visualization team at the Barcelona Supercomputing Center (BSC-CNS) for the insightful conversations and suggestions. RS acknowledges partial support from the European Union’s Horizon 2020 research and innovation program under grant agreement No. 869764 (GoGreenRoutes). RS acknowledges partial support from the Severo Ochoa Mobility Program at BSC-CNS. The funders had no role in study design, data collection, and analysis, decision to publish, or preparation of the manuscript.

\section{Contributions}
AB performed the data collection, processed and cleaned the data, built the database and the back end of the interactive web platform, designed and performed the analysis, and drafted the manuscript. RS designed and supervised the study, revised the manuscript, and built the front end for the interactive web platform. AB performed most of the work while at the Barcelona Supercomputing Center (BSC-CNS).


\clearpage
\setcounter{figure}{0}
\setcounter{table}{0}
\setcounter{equation}{0}
\makeatletter
\renewcommand{\thefigure}{S\arabic{figure}}
\renewcommand{\theequation}{S\arabic{equation}}
\renewcommand{\thetable}{S\arabic{table}}

\setcounter{secnumdepth}{2}

\widetext
\begin{center}
	\textbf{\large Supplemental Material:\\ On the need to move from a single indicator to a multi-dimensional framework to measure accessibility to urban green}
\end{center}

\section{Data cleansing and processing} \label{SISEC_DataCleaningProcessing}
This section provides detailed information on data cleansing and processing. In what follows, the terms \textit{urban center} (UC) and \textit{city} are used interchangeably.

\subsection{Definition of urban center and sample of cities} \label{SISEC_UCDB}
UCs were defined according to the definition in the Urban Centre Database of the Global Human Settlement 2015, revised version R2019A (GHS-UCDB) \cite{florczyk2019}. As such, they are identified by specific cut-off values on the resident population and the built-up surface share of a regular 1x1 km global grid and do not coincide with administrative entities.
The database records more than 13,000 urban centers. Out of these, we only retained the most populated 50 UCs per country (three-code isocode), provided that they have at least 100,000 inhabitants (corresponding to roughly 2,600 UCs).
To ensure only cities with sufficient quality of OpenStreetMap (OSM) data are included in the study, we filtered the sample according to the following two rules: 
\begin{enumerate*}
\item we excluded from the sample, all UCs with less than 20 distinct public green areas recorded in OSM (about 600 cities were excluded, 1963 cities were left in the sample). To this scope, public green areas were defined according to the $key:value$ pairs in Table \ref{tab:OSM_natural_element_key_val_pairs}
\item we further excluded 923 urban centers, according to the data validation step described in Section \ref{SISEC_dataValidation} of the SM.
\end{enumerate*} 
The final sample for the study comprised 1,040 urban centers across 145 countries. 

\subsection{Population data}
Population data were extracted from the population grid of the Global Human Settlement 2015 (revised version 2019A) \cite{ghs_pop} at a 9-arcseconds resolution. The data consist of residential population estimates for the year 2015, disaggregated from census or administrative units to grid cells and informed by the distribution and density of built-up as mapped in the corresponding Global Human Settlement Layer (GHSL) global layer. The data were downloaded as a global \textit{.tiff} file. For each urban center, the data were then extracted and processed using the \textit{rioxarray} package in Python (\href{https://corteva.github.io/rioxarray/stable/}{https://corteva.github.io/rioxarray/stable/}), masking the global file with the boundary of the city (see \ref{SISEC_UCDB}) enlarged with a three km buffer and subsequently the clipped raster was loaded into a PostGIS database. Throughout the study, the population grid was used as \textit{base-grid}, i.e. cells of the population grid represent the smallest unit of analysis. 

\subsection{Public green areas from OpenStreetMap} \label{SISEC:OSM_Green}
The OpenStreetMap (OSM) data dump for each continent was downloaded from the GeoFabrik Download Service (\href{http://www.geofabrik.de/}{http://www.geofabrik.de/}) on May 2022. For each urban center, the data were processed with a three-step pipeline:
\begin{enumerate*}
    \item Extraction of local osm.pbf dumps;
    \item Extraction of public green areas from OSM;
    \item Remapping of public green area to the base-grid.
\end{enumerate*}
Each step is described in detail below.

\paragraph*{Extraction of local osm.pbf dumps}
For each UC, the first step consisted of the generation of the geographical extract of the OSM data from the OSM dump of the corresponding continent. To ensure that locations at the boundary of the study area are not biased, the perimeter of the urban area defined in the GHS-UCDB \cite{florczyk2019} was buffered with a three km radius. The extraction was performed using the \textit{smart strategy} of the method \textit{extract} of the Osmium Library. The \textit{smart strategy} runs in three passes and extracts: 
\begin{enumerate*} 
\item all nodes inside the region and all ways referencing those nodes as well as all nodes referenced by those ways; 
\item all relations referenced by nodes inside the region or ways already included and, recursively, their parent relations; 
\item all nodes and ways (and the nodes they reference) referenced by relations tagged \textit{type=multipolygon} directly referencing any nodes in the region or ways referencing nodes in the region. 
\end{enumerate*} 
The default configuration was used for all other options of the method. 

\paragraph*{Extraction of public green areas from OSM}
For each UC, the second step consisted of the extraction of public green areas from the local osm.pbf file. In this study, we defined as pubic green areas all those OSM elements (closed ways and relations) with the $key:value$ pairs defined in Table \ref{tab:OSM_natural_element_key_val_pairs}.
The extraction was performed using the Python bindings of the osmium-tool (\href{https://osmcode.org/osmium-tool/}{https://osmcode.org/osmium-tool/}). The geometries of relations were reconstructed from their members using the approach described \href{https://wiki.openstreetmap.org/wiki/Relation:multipolygon/Algorithm.}{here}.
Elements with $key:value$ pairs equal to \textit{access:no} or \textit{access:private} were excluded from the retrieved database. In the final step, we classified green OSM elements into three categories (Parks, Forests, Grass) according to the mapping provided in  Table \ref{tab:OSM_natural_element_key_val_pairs}. The obtained green features were then loaded into the PostGIS database.

\begin{table}[t!]
\centering
\begin{tabular}{lll|lll|lll}
\hline
\textbf{Category}      & \textbf{Key} & \textbf{Value} & \textbf{Category}        & \textbf{Key} & \textbf{Value} & \textbf{Category}      & \textbf{Key} & \textbf{Value} \\ \hline
\multirow{4}{*}{Parks} & leisure      & park           & \multirow{4}{*}{Forests} & landuse      & forest         & \multirow{4}{*}{Grass} & landuse      & grass          \\
 & leisure & garden             &  & natural & wood &  & landuse & meadow    \\
 & landuse & recreation\_ground &  &         &      &  & natural & grassland \\
 & landuse & village\_green     &  &         &      &  & natural & shrubbery \\ \hline
\end{tabular}
\caption{OSM $key:value$ pairs used for the identification of public green areas}
\label{tab:OSM_natural_element_key_val_pairs}
\end{table}

\paragraph*{Remapping of public green area to the base grid}

For each UC, the green OSM elements were then mapped to the base grid used for analysis. To each cell in the grid and for each combination of green elements ({parks}, {forests}, {grass}, {parks \& forests}, etc.) intersecting the cell, we associate information on:
\begin{enumerate*}
\item size of the intersection between the green element and the cell;
\item the overall size of the green element. 
\end{enumerate*}
All values were expressed in hectares and rounded to the second decimal point so that the minimum resolution is a continuous green space of at least 100 square meters.
It should be noted that green OSM elements might partially overlap with each other (for instance the same area could be mapped as $leisure:parks$ and $landuse:grass$). To ensure the same area is not computed multiple times and that continuous areas of green (for instance adjacent forests and grasslands) are properly identified, overlapping areas were dissolved together before proceeding with the remapping. 

\subsection{Green coverage from the World Cover data 2020 of the European Space Agency} \label{Supp:ESA_Green}
The World Cover data 2020 \cite{zanaga_daniele_2021_5571936} from the European Space Agency was used in this study for the computation of the exposure index as well as reference data to identify cities with sufficient good quality of OpenStreetMap data, as described in the \ref{SISEC_dataValidation}. The data provides global land cover maps for 2020 at 10 m resolution based on Copernicus Sentinel-1 and Sentinel-2 data. Out of all cover classes, classes \textit{10 - Tree cover}, \textit{20 - Shrubland} and \textit{30 - Grassland} were defined as \textit{green coverage}. For each city, the data were extracted from the global .tiff, by clipping the worldwide file with the boundary of the UC buffered with
a 3 kilometers radius following the procedure
described \href{https://esa-worldcover.org/en/data-access}{here} and subsequently loaded into the PostGIS database. The data were then remapped to the base grid of each city following the approach described in \ref{SISEC:OSM_Green}-\textit{Remapping of public green area to the base grid}.

\subsection{Computation of walking distance matrices}
\begin{figure}[b]
\includegraphics[width=\linewidth]{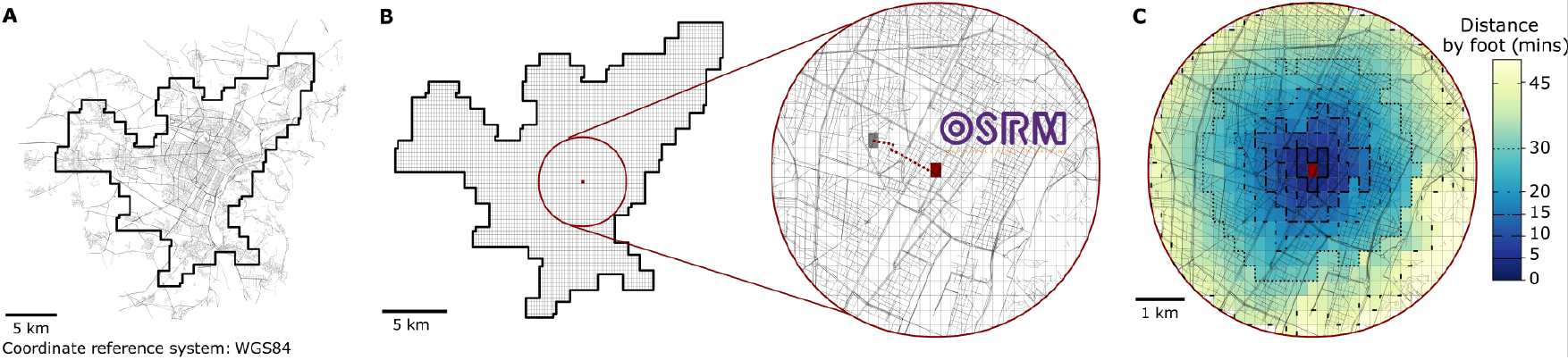}
\caption{\textbf{Computation of walking distance matrices. An example for the city of Turin, Italy.} A and B provide a schematic representation of the pipeline for the computation of the walking distances, for the city of Turin, Italy. A depicts the street-network for the buffered region (GHS-UCDB boundary buffered with a 3 km radius) was extracted from the local OSM dump and the city boundary (solid black line). B depicts the base grid of the city of Turin. For each cell (depicted in red in the example), the walking distance between the centroid of the cell itself and the centroids of all cells within a 3 km buffered region (red circle) was computed using the local routing engine Open Source Routing Machine. C provides an example of the walking distances computed using the described approach. The origin cell was depicted in red. Solid/Dashed/Dotted lines indicate several isochrones for the cell of interest.}
\label{fig_SI:DistanceMatrices}
\end{figure}

In this study, the problem of computing the walking distances between residential areas and green spaces in each urban area was simplified to the computation of walking distances between the centroids of the base grid, i.e. of computing a walking origin-destination matrix. To this scope, we locally installed the routing engine Open Source Routing Machine (OSRM) \cite{luxen-vetter-2011} and used street-network data from the local OSM dumps (see \ref{SISEC:OSM_Green}-\textit{Extraction of local osm.pbf dumps}). A schematic representation of the adopted strategy for the city of Turin, Italy, is provided in Fig. \ref{fig_SI:DistanceMatrices}.

In particular, for each UC $c$, we used a four-step pipeline.
\begin{enumerate}
    \item  From all cells in the extended grid of the city $\mathcal{M}$, define the set $\mathcal{Z}$ (with members: $z$) to be the subset of $\mathcal{M}$ with non-null population.
    \item Characterize each cell $z$ in  $\mathcal{Z}$ and each cell $m$ in $\mathcal{M}$ with the lat-long coordinates of their centroids. 
    \item For each $z$ in $Z$:
    \begin{enumerate*}
         \item Identify the subset of cells in $\mathcal{M}$ within a geodesic buffer of 3 kilometers. Call this set $\mathcal{M'}_{z}$. It should be noted that for computational reasons, we did not compute the full distance matrix, but restricted the computation to cells that are no more than 3km away (geodesic distance) from each other. Given that the focus of this work is on walking distance, this simplification has no impact on our results. 
        \item Use the matrix routing engine OSRM (with default configuration parameters for the \textit{foot} profile) to compute the walking distance in minutes between the centroid of z and the centroids of all  $m'$ in $\mathcal{M'}_{z}$.
   \end{enumerate*}
\end{enumerate}

As the matrix approach of OSRM may provide incorrect results if the starting/ending point is not sufficiently close to a walkable street, we used a heuristic approach to identify potentially problematic starting and ending cells, by intersecting ex-ante the base-grid with the walkable street network (defined mimicking the parameters of the \textit{foot} profile in OSRM). If a cell presents no intersecting street, then it is assumed not to be accessible and all distances are set to NA.
As a term of comparison, we additionally generated distance matrices based on geodesic distance, by replacing point 4.2 above with a simple computation of the geodesic distance between the centroids of the two cells.

\section{Validation of the sample of cities} \label{SISEC_dataValidation}
 The quality of OSM data is under constant scrutiny by researchers relying on this data source for their studies. On one side, this led to the development of specific web services to visually compare OSM data with other mapping systems - such as the \href{https://mc.bbbike.org/mc/}{OSM Map Compare tool} provided by  BBBike and Geofabrik (\href{http://www.geofabrik.de/}{http://www.geofabrik.de/}). On the other side, several methods were developed by the scientific community to quantify the quality of the OSM data, ranging from methods of intrinsic evaluation (where the metrics are defined using OSM data) to methods based on the comparison with external data sources (reference data). More generally, as the suitability of OSM data for a particular application depends largely on the specific problem being tackled, researchers cannot rely on a pre-defined list of areas, but should rather evaluate the quality of the data in the context of the specific problem they face.
Here, we propose an approach to validate the sample of cities to include in our study, which is based on the comparison with the World Cover data of the European Space Agency \cite{zanaga_daniele_2021_5571936}.
The pipeline for the data validation comprised two phases:
\begin{enumerate}
    \item Definition of a Quality Score of green elements, for each city in the sample;
    \item Identification of a set of cities with similarly-high data quality using k-means clustering.
\end{enumerate}
Each step is defined in detail below.

\paragraph{Definition of a Quality Score of green elements, for each city in the sample}

\begin{figure}[t]
\includegraphics[width=\linewidth]{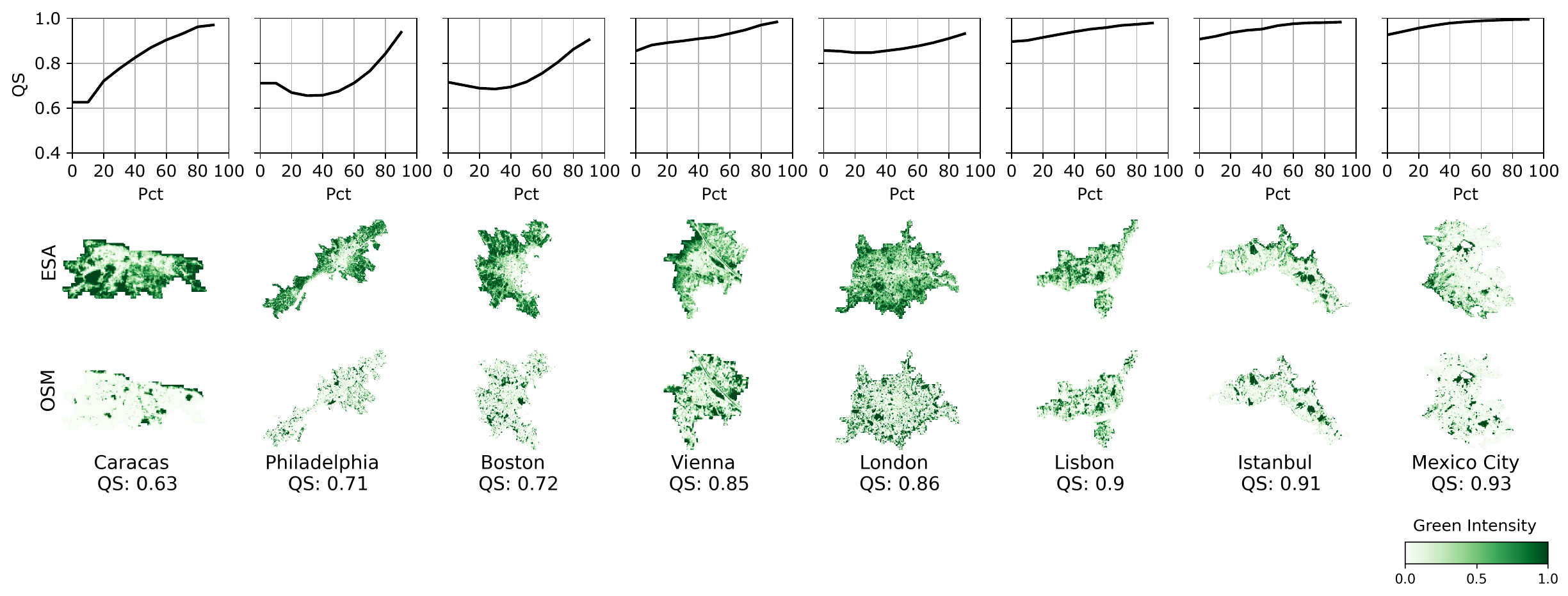}
\caption{\textbf{Green intensity images and quality score: the example of eight cities.} The chart displays the green intensity images generated using data from the European Spatial Agency World Cover 2020 data set (middle row) and data from OpenStreetMap (bottom row), for the cities of Caracas (Venezuela), Philadelphia (United States of America), Boston (United States of America), Vienna (Austria), London (United Kingdom), Lisbon (Portugal), Istanbul (Turkey) and Mexico City (Mexico). For each city, the top row displays the change in the Quality Score obtained by removing from the images cells/pixels of the progressively increasing population [i.e. the bottom 10\% (20\%, 30\%, etc) of the cells].}
\label{fig_SI:DataValidtion_1proc}
\end{figure}

For each city in our sample, we computed the OSM green data Quality Score (QS), based on the comparison of green features extracted from OpenStreetMap and green features observable in the World Cover data 2020 of the Europen Satellite Agency \cite{zanaga_daniele_2021_5571936}, here used as an external reference dataset. 
To this scope, we extended the list of green features in OpenStreetMap to include features that -whilst not classifiable as public green areas - might appear as green cover in the satellite dataset. A list of additional OSM tags retained to this scope is provided in Table \ref{tab:OSM_green_addition}. Only ways and relations were extracted and the choice of tags was based on a manual assessment of randomly picked cities. The data were processed and remapped to the base grid following the approach described in \ref{SISEC:OSM_Green}.

Borrowing techniques from the digital image processing domain, the QS was defined to be proportional to the mean squared error between a green intensity image generated using OSM data and an equivalent picture generated using land cover data from the European Spatial Agency. 
In particular, for both data sources and for each city, we first projected the green information on a nine-arc-sec resolution grid and characterized each cell with a green intensity value corresponding to the proportion of the cell that is covered by a green feature (the values range between 0 and 1). By treating each cell in the grid as a pixel, we then generated two images of green intensity, one using data from OpenStreetMap and the other using data on the World Cover from the European Space Agency. For each city in the sample, we then computed the mean squared error among the two images as:
\begin{equation}
MSE_{c}=\dfrac{1}{N}\sum_{i \in I} (GI_{OSM, i} - GI_{ESA, i})^{2}
\end{equation}
where $GI_{data, i}$ is the green intensity value of pixel $i$ from the image generated using data from $data$, $I$ is the set of pixels (cells) composing the image and $N$ is the cardinality of $I$.
The $MSE$ is bounded between 0 (identical green intensity) and 1 (complementary green intensity).
For each city $c$, the QS was then defined as 
\begin{equation}
QS_{c}=1-MSE_{c}
\end{equation}

A visual example for eight cities is provided in Fig. \ref{fig_SI:DataValidtion_1proc}.

\begin{table}[b]
\centering

\begin{tabular}{ll|ll|ll|ll}
\hline
\textbf{Key} & \textbf{Value}    & \textbf{Key} & \textbf{Value}     & \textbf{Key} & \textbf{Value}         & \textbf{Key} & \textbf{Value} \\ \hline
 amenity & grave\_yard     landuse    & landuse & orchard    & landuse & farmyard      &  natural & heath \\
  landuse & cemetery &  landuse  & flowerbed &  landuse  & construction  &         natural     & tree\_row      \\
 amenity & school              & landuse & greenfield &  leisure & pitch         & natural & fell  \\
 amenity            & kindergarten      &      landuse        & farmland  &        leisure      & horse\_riding &        natural      & wetland        \\
  amenity & education           &  landuse & vineyard   & natural & scrub         &  &       \\
landuse & allotments  &  landuse & farm       & natural & moor &  &       \\ \hline
\end{tabular}

\caption{OSM $key:value$ pairs used for the identification of the additional sources of green coverage}
\label{tab:OSM_green_addition}
\end{table}

\paragraph{Identification of a set of cities with similarly-high data quality using k-means clustering}
\begin{figure}[t]
\includegraphics[width=0.95\linewidth]{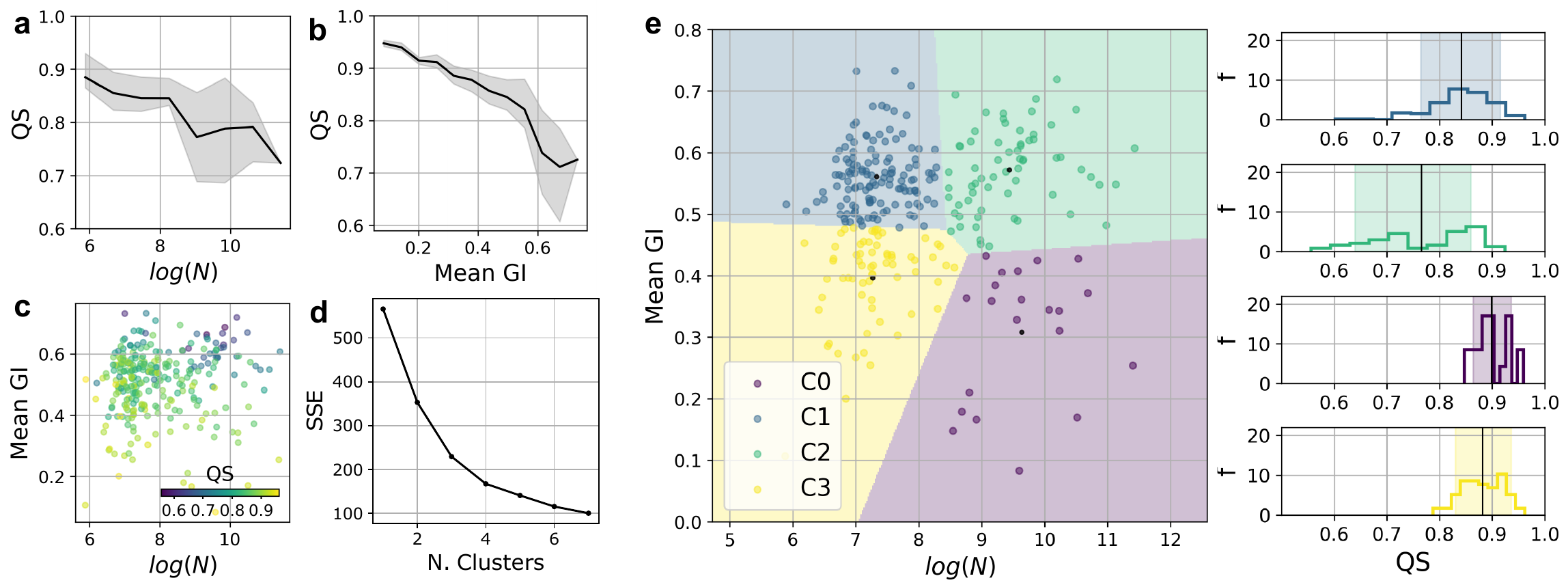}
\caption{\textbf{Validation of the sample of cities} a) displays the Average (solid line) and IQR (shaded area) Quality Scores of cities in the \textit{reference list} as a function of the logarithm of the number of pixels/cells ($log(N)$). b) displays the average (solid line) and IQR (shaded area) Quality Scores of cities in the \textit{reference list} as a function of the Mean GI, i.e. the average value of green intensity for pixels within the boundary of the city. c) is a scatter plot of the $log(N)$ and \textit{Mean GI}. Each point represents a city in the \textit{reference list}. The color represents the Quality Score. d represents. d) represents the k-means inertia for different numbers of clusters, obtained when clustering cities in the \textit{reference list} based on their tuple of values ($log(N)$, \textit{Mean GI}). A z-scores transformation was performed before clustering the data. e) is a scatter plot of obtained clusters and corresponding regions s identified by the corresponding decision boundaries. The insects on the right depict the distribution of the QS of cities in the \textit{reference list} (histogram) and the admittable region (shaded area) for each cluster (C=: purple, C1: blue, C2: green, C3: yellow).}
\label{fig_SI:DataValidtion_2proc}
\end{figure}

The second step consisted of the identification of a set of cities with sufficiently-high and homogeneous quality. To this scope, we proceeded as outlined below:
\begin{enumerate}
    \item First, we selected a sample of cities (\textit{reference list}) from countries where OSM data are generally used for academic research. While we do not evaluate specifically the quality of the mapping for these cities, our scope was to identify -- among the remaining urban centers -- those with comparable data quality, to ensure the comparison proposed in the study is valid. We included in the \textit{reference list}, all those cities in our initial sample (see \ref{SISEC_UCDB}) located in Italy, France, Netherlands, Germany, Austria, United Kingdom, Luxembourg, Belgium, the United States, and Canada.  
    \item For cities in the \textit{reference list}, we then studied the association between the QS and two city-level metrics: the size (captured by the natural logarithm of the number of pixels in the image, or $log(N)$) and the average green intensity (measured taking the average of the green intensity of the image generated using the World Cover data from the European Space Agency across all pixels, or $\textit{Mean GI}=\dfrac{1}{N}\sum_{i \in I}GI_{ESA, i}$). As depicted in panels a, b, and c of Fig. \ref{fig_SI:DataValidtion_2proc}, the QS decreases with the size and the average greenness of the city. I.e. larger and/or greener cities have a lower QS than smaller and less green ones. It should be noted that this negative association depends on the definition of the QS rather than implying a difference in the quality of the mapping. Indeed: 
    \begin{itemize}
    \item the QS of cities with low green coverage is artificially inflated by the absence of features to be mapped. This is the case because, the lower the number of features to be mapped, the lower the information that the QS provides. A powerful example is the extreme case of cities with no green coverage at all. By construction the QS would be equal to 1 in these cases, however, a QS of one does not guarantee a generally good quality of OSM data. 
    \item for geometrical reasons, larger cities have more peripheral pixels, with lower population density. The positive association between the population density and the QS (see top row of Fig. \ref{fig_SI:DataValidtion_1proc}) determines the lower QS observed for large cities. 
    \end{itemize}
    \item Due to the negative association between the size of the city and its average greenness and the QS, the use of a single cut-off value to identify cities to be included in our study might result in the inclusion of some urban centers with insufficient data quality (for instance if the city has poor green coverage) or the exclusion of others with sufficient data quality (for example large cities). By k-means clustering cities in the \textit{reference list} based on the size and the average greenness, we identified four regions of tuples of values ($log(N)$, \textit{Mean GI}) (Panel e in Fig. \ref{fig_SI:DataValidtion_2proc}) and defined an admittable range of QS values as values between the $10th$ and the $90th$ percentile of the QS of cities in the \textit{reference list} belonging to the specific cluster (shaded areas in the insects of panel e of Fig. \ref{fig_SI:DataValidtion_2proc}). Standard preprocessing was performed for the k-means clustering.
    \item The final sample of cities was then defined as the union of cities in the \textit{reference list} and cities outside the \textit{reference list} whose QS fell within the admittable interval for their tuple of ($log(N)$, \textit{Mean GI}) values. 
\end{enumerate}
The final sample for the study comprises 1,040 urban centers.

\newpage
\section{Stability of the exposure and per-person indicators}

\begin{figure}[h]
\includegraphics[width=\linewidth]{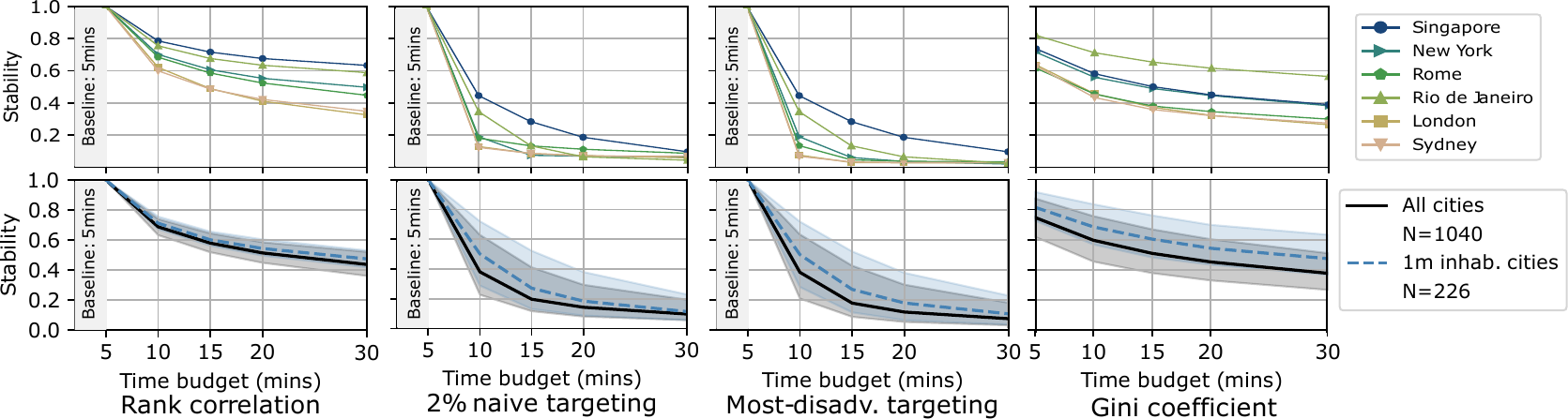}
\caption{\textbf{Stability of the exposure indicator to the time-budget} The top row depicts the level of stability of the exposure index to different time-budgets and according to different stability metrics, for six cities across all continents. For the \textit{Rank correlation}, \textit{2\% naive targeting} and the \textit{Most-disadvantaged targeting}, the comparison is provided with respect to the parametrization with time-budget equal to 5 mins. For the \textit{Gini index}, the chart reports the value of the index under several parametrizations. The bottom row reports the median value (solid line) and the IQR (shaded area) of the stability metrics for all cities in our sample (black) and for cities with more than 1 million inhabitants (blue). A formal definition of each stability metric is provided in the Materials and Methods.}
\label{SI_EXP_stab}
\end{figure}

\newpage

\begin{figure}[h]
\includegraphics[width=\linewidth]{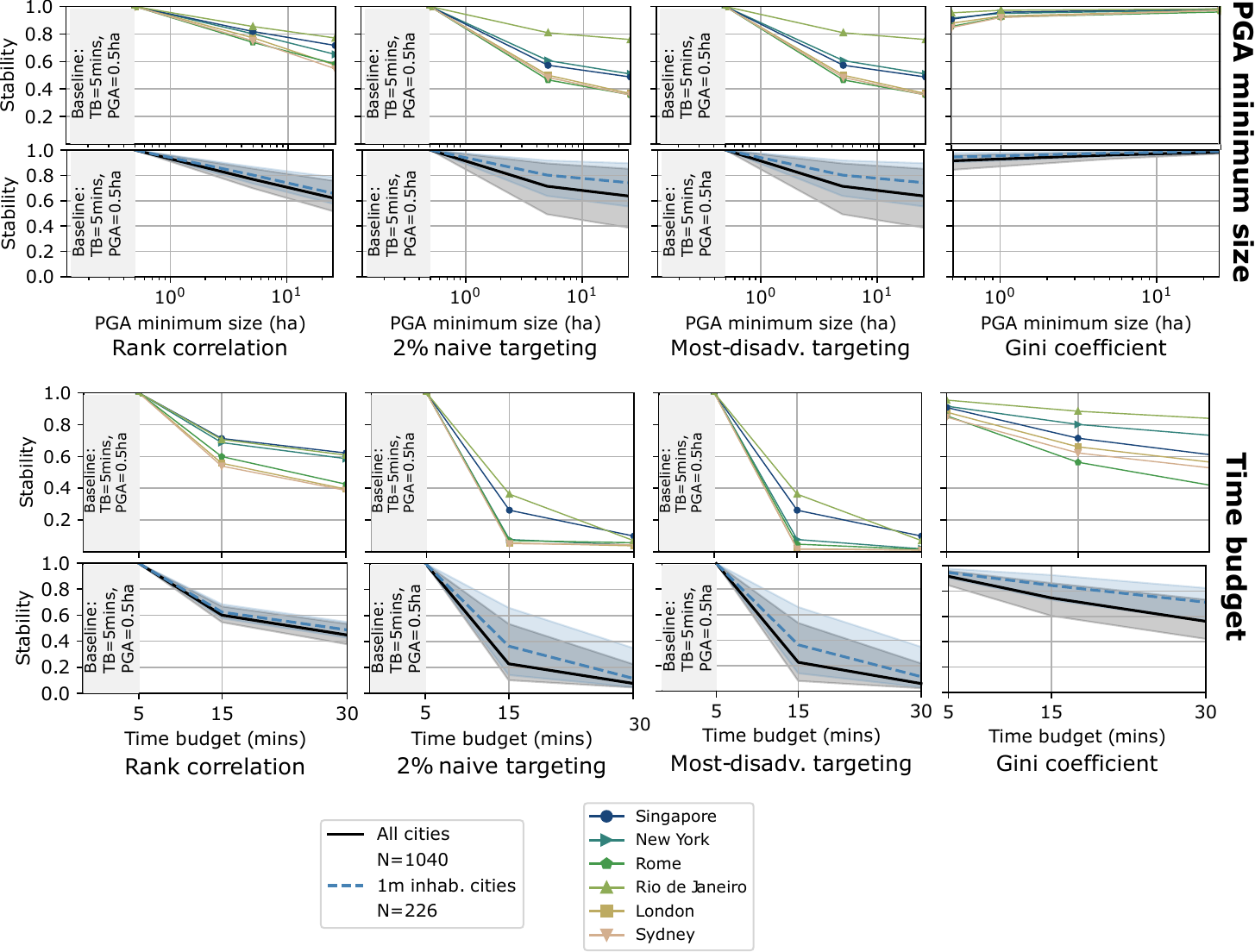}
\caption{\textbf{Stability of the per-person indicator to the minimum size of the PGA and the time-budget}. The panel depicts the level of stability of the per-person index to several minimum PGA sizes (top sub-panel) and time-budgets (bottom sub-panel). 
For both sub-panels: the top row displays the level of stability of the per-person index to the parameter and according to four stability metrics, for six cities across all continents . For the \textit{Rank correlation}, \textit{2\% naive targeting} and the \textit{Most-disadvantaged targeting}, the comparison is provided with respect to the parametrization with time-budget equal to 5 mins and minimum PGA size equal to 0.5 ha. For the \textit{Gini index}, the chart reports the value of the index under several parametrizations; the bottom row reports the median value (solid line) and the IQR (shaded area) of the stability metrics for all cities in our sample (black) and for cities with more than 1 million inhabitants (blue). A formal definition of each stability metric is provided in the Materials and Methods. }
\label{SI_PP_stab}
\end{figure}

\newpage

\section{Sensitivity of the stability metrics to the targeting strategies}

\begin{figure}[h]
\includegraphics[width=0.8\linewidth]{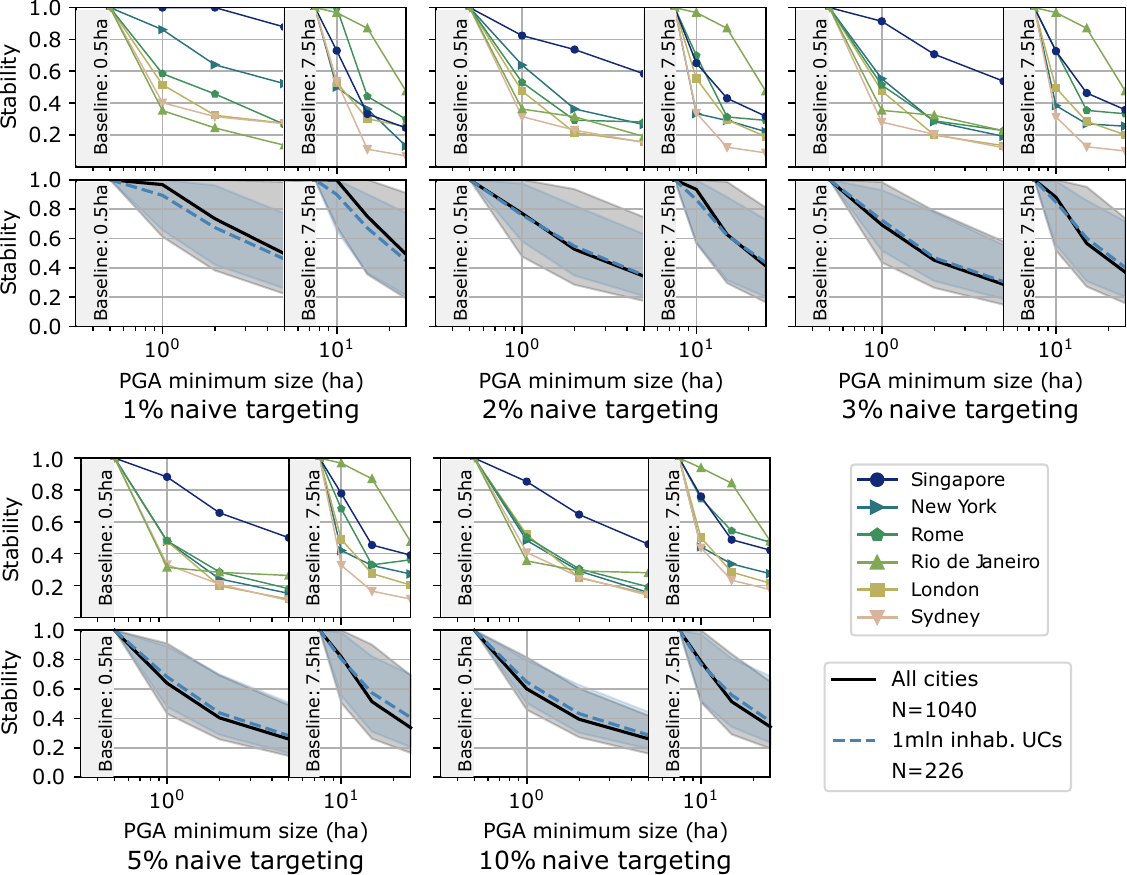}
\caption{\textbf{Stability of the minimum distance indicator to the minimum size of the PGA:  $y\%$ naive targeting strategies} The panel depicts the stability of the minimum distance indicator to the minimum PGA size under several naive targeting strategies [1\%, 2\%, 3\%, 5\%,10\%], for six cities, the median (solid line) and IQR (shaded area) across all cities (black) and the median (solid line) and IQR (shaded area) for cities with more than 1 million inhabitants (blue)}
\label{SI_MD_sensitivity_naive}
\end{figure}

\newpage

\begin{figure}[h]
\includegraphics[width=\linewidth]{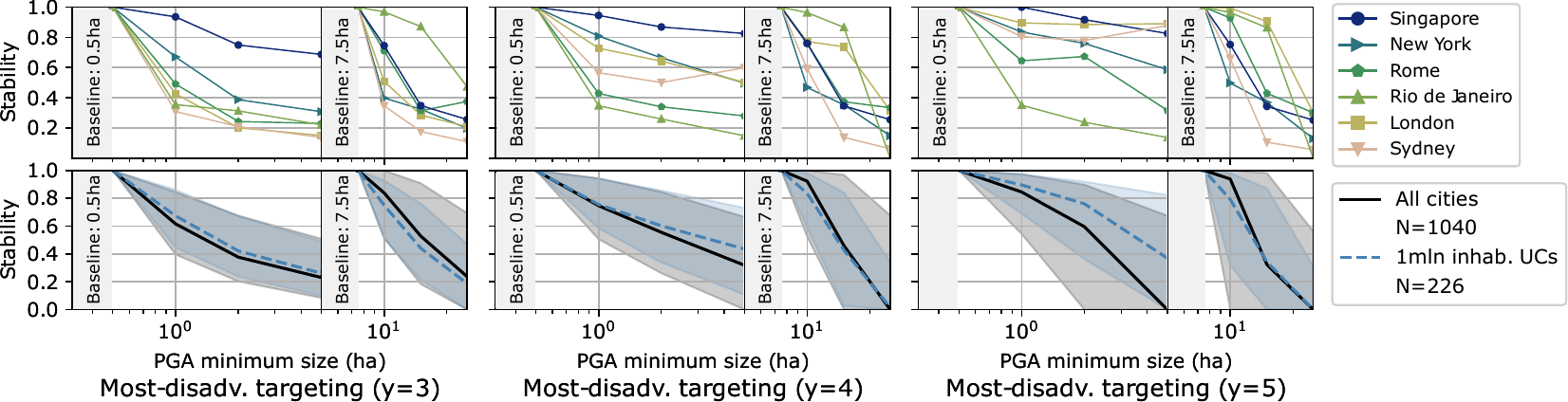}
\caption{\textbf{Stability of the minimum distance indicator to the minimum size of the PGA: $y$-times most-disadvantaged targeting strategies} The panel depicts the stability of the minimum distance indicator to the minimum PGA size under several most-disadvantaged targeting strategies [3-times worse than mean citizen, 4-times worse than mean citizen and 5 -times worse than mean citizen], for six cities, the median (solid line) and IQR (shaded area) across all cities (black) and the median (solid line) and IQR (shaded area) for cities with more than 1 million inhabitants (blue)}
\label{SI_MD_sensitivity_most_disadv}
\end{figure}

\newpage

\begin{figure}[h]
\includegraphics[width=0.8\linewidth]{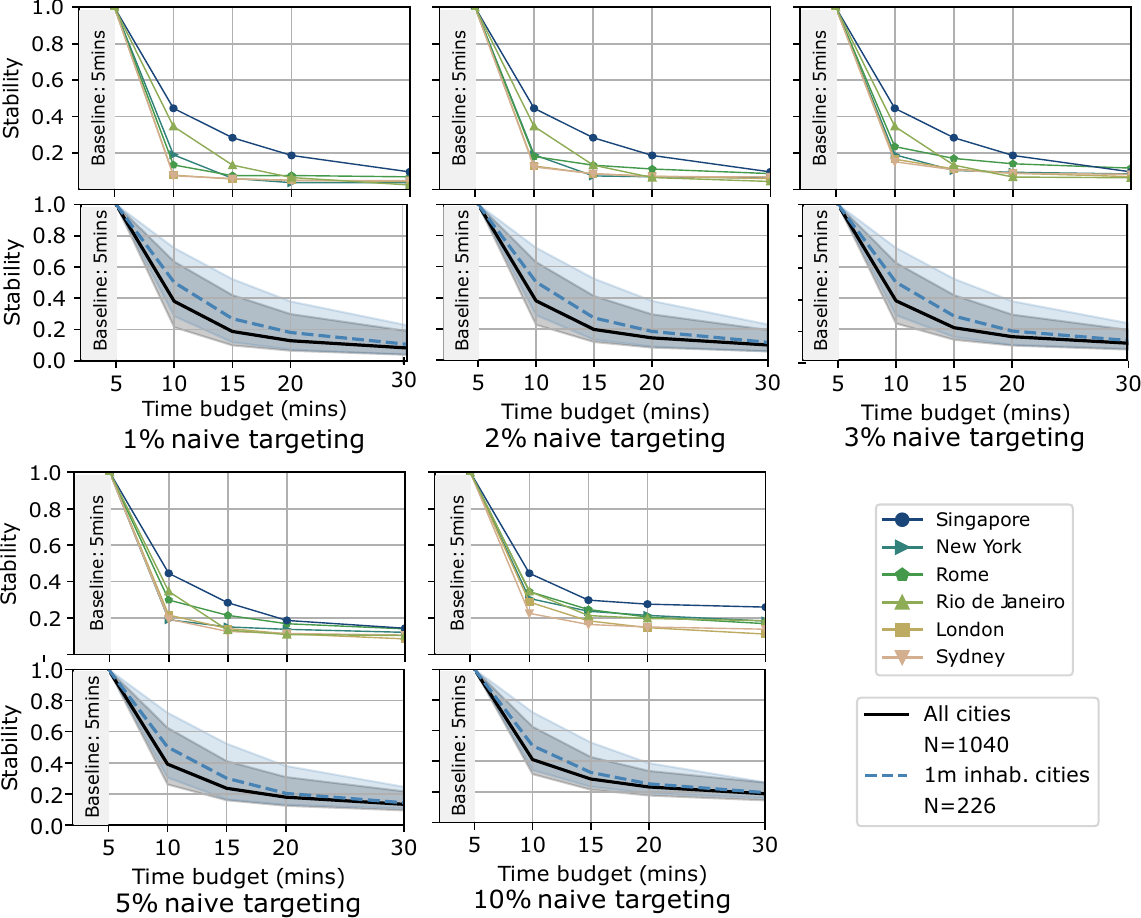}
\caption{\textbf{Stability of the exposure indicator to the time budget: $y\%$ naive targeting strategies} The panel depicts the stability of the exposure indicator to the time budget under several naive targeting strategies [1\%, 2\%, 3\%, 5\%,10\%], for six cities, the median (solid line) and IQR (shaded area) across all cities (black) and the median (solid line) and IQR (shaded area) for cities with more than 1 million inhabitants (blue)}
\label{SI_EXP_sensitivity_naive}
\end{figure}

\newpage

\begin{figure}[h]
\includegraphics[width=0.8\linewidth]{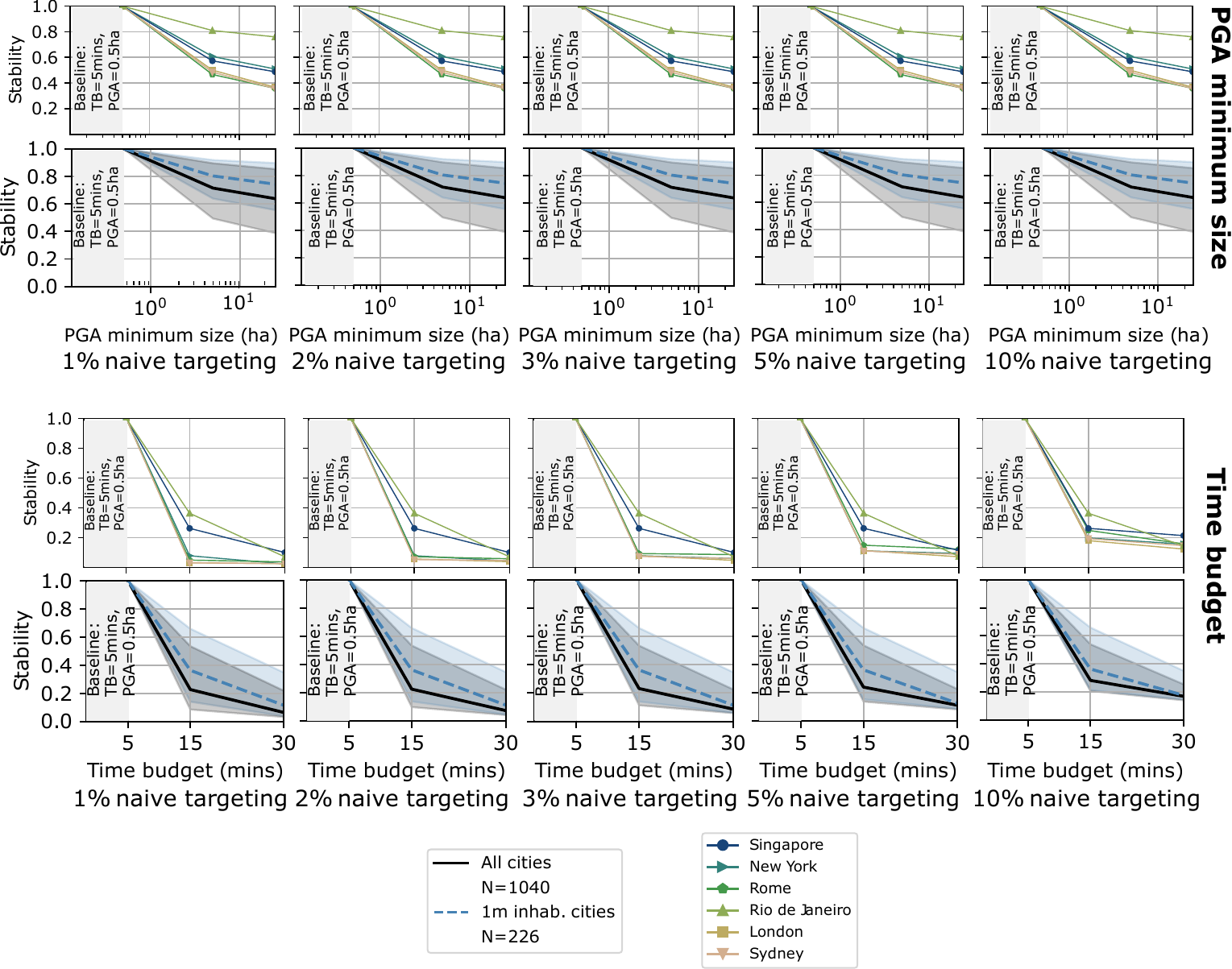}
\caption{\textbf{Stability of the per-person indicator to the minimum size of the PGA and time budget: $y\%$ naive targeting strategies} The panel depicts the level of stability of the per-person index to several minimum PGA sizes (top sub-panel) and time-budgets (bottom sub-panel) under several naive targeting strategies [1\%, 2\%, 3\%, 5\%,10\%], for six cities, the median (solid line) and IQR (shaded area) across all cities (black) and the median (solid line) and IQR (shaded area) for cities with more than 1 million inhabitants (blue). }
\label{SI_PP_sensitivity_naive}
\end{figure}

\end{document}